\documentclass[10pt,letterpaper]{article}
\usepackage[margin=1in]{geometry}

\usepackage{times}
\usepackage{soul}
\usepackage{url}
\usepackage[hidelinks]{hyperref}
\usepackage[utf8]{inputenc}
\usepackage[small]{caption}
\usepackage{graphicx}
\usepackage{amsmath}
\usepackage{amsthm}
\usepackage{booktabs}
\usepackage{algorithm}
\usepackage{algorithmic}
\usepackage{url}
\usepackage{xspace}
\usepackage{booktabs}
\usepackage{amsmath, amssymb}
\usepackage{mathtools}
\usepackage{breqn}
\usepackage{enumitem}

\usepackage{braket,amsfonts,amsopn}
\usepackage{algorithm}
\usepackage{algorithmic}
\usepackage{xfrac}
\usepackage{color}

\usepackage{graphicx}
\usepackage{subcaption}
\usepackage{caption}
\usepackage{dsfont}

\newcommand{\smallsection}[1]{\vspace{0.5mm}{\noindent {\bf{\underline{\smash{#1}}}}}}
\newtheorem{lma}{\textbf{Lemma}}
\newtheorem{thm}{\textbf{Theorem}}
\newtheorem{pro}{\textbf{Problem}}
\newtheorem{dfn}{\textbf{Definition}}

\newcommand{\blue}[1]{\textcolor{blue}{#1}}

\newcommand{\method}{\textsc{HashNWalk}\xspace}
\newcommand{\midas}{\textsc{Midas}\xspace}
\newcommand{\sedanspot}{\textsc{SedanSpot}\xspace}
\newcommand{\ffade}{\textsc{F-FADE}\xspace}
\newcommand{\lsh}{\textsc{LSH}\xspace}

\newcommand{\syntheticU}{\texttt{SemiU}\xspace}
\newcommand{\syntheticB}{\texttt{SemiB}\xspace}
\newcommand{\real}{\texttt{Transaction}\xspace}

\title{HashNWalk: Hash and Random Walk Based Anomaly Detection in Hyperedge Streams}

\author{
Geon Lee
\and
Minyoung Choe \and
Kijung Shin
}

\date{\normalsize Kim Jaechul Graduate School of AI, KAIST, Seoul, South Korea \\
\{geonlee0325, minyoung.choe, kijungs\}@kaist.ac.kr}

\begin{document}

\maketitle

\begin{abstract}
Sequences of group interactions, such as emails, online discussions, and co-authorships, are ubiquitous; and they are naturally represented as a stream of hyperedges. 
Despite their broad potential applications, anomaly detection in hypergraphs (i.e., sets of hyperedges) has received surprisingly little attention, compared to that in graphs. 
While it is tempting to reduce hypergraphs to graphs and apply existing graph-based methods, according to our experiments, taking higher-order structures of hypergraphs into consideration is worthwhile.
We propose \method, an incremental algorithm that detects anomalies in a stream of hyperedges. It maintains and updates a constant-size summary of the structural and temporal information about the stream. Using the summary, which is the form of a proximity matrix, \method measures the anomalousness of each new hyperedge as it appears. \method is \textbf{(a) Fast:} it processes each hyperedge in near real-time and billions of hyperedges within a few hours, \textbf{(b) Space Efficient}: the size of the maintained summary is a predefined constant, \textbf{(c) Effective:} it successfully detects anomalous hyperedges in real-world hypergraphs.
\end{abstract}
    
    \section{Introduction}
	\label{sec:intro}
	\begin{figure*}[t]
	\vspace{-4mm}
	\centering
    \begin{subfigure}[b]{.31\textwidth}
          \centering
          \includegraphics[width=0.985\columnwidth]{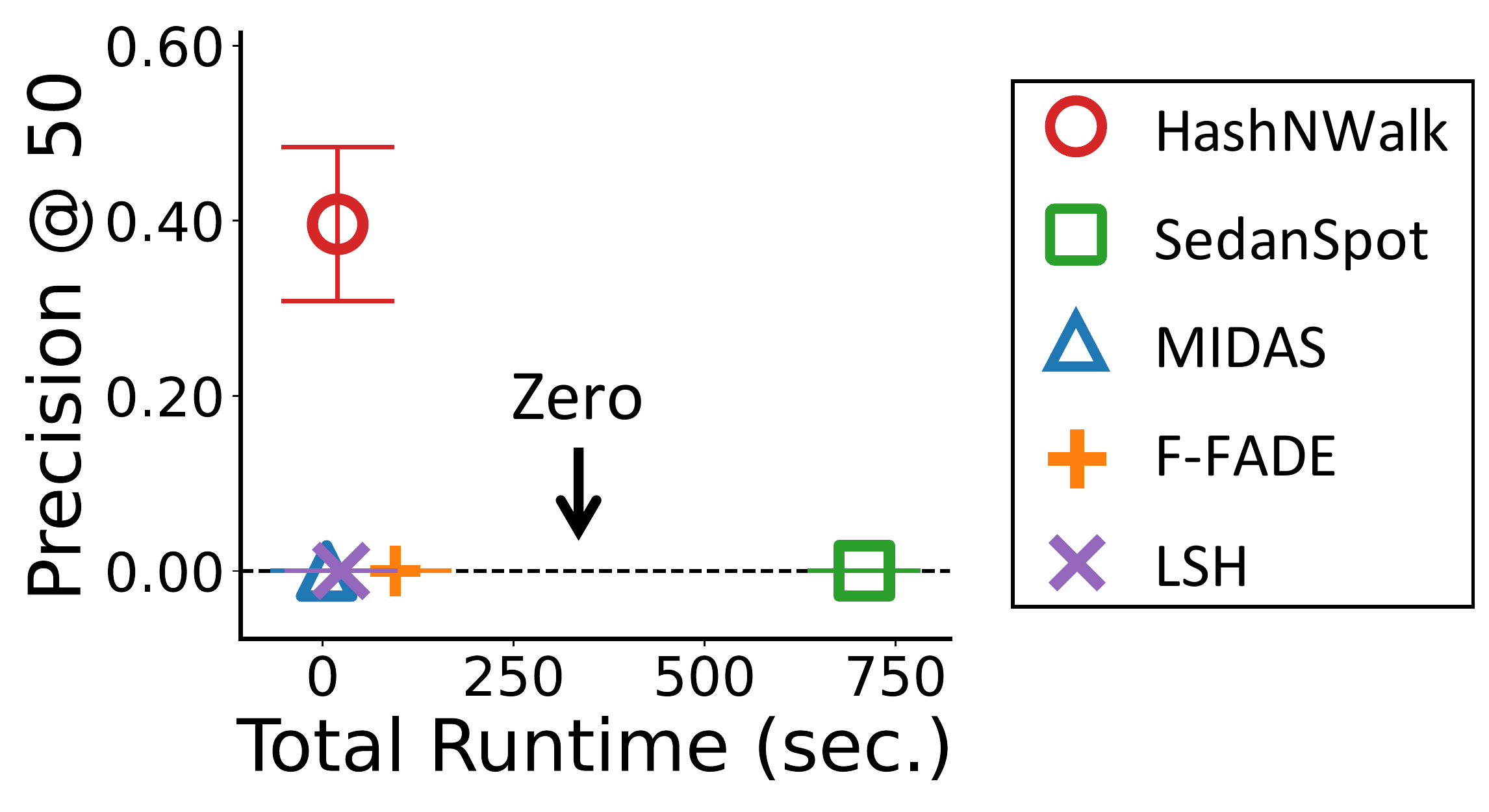}
		  \vspace{-16pt}
          \caption{Speed and preciseness}
          \label{fig:crown:accuracy}
    \end{subfigure}
    \hspace{0pt}
    \begin{subfigure}[b]{.215\textwidth}
          \centering
          \includegraphics[width=0.950\columnwidth]{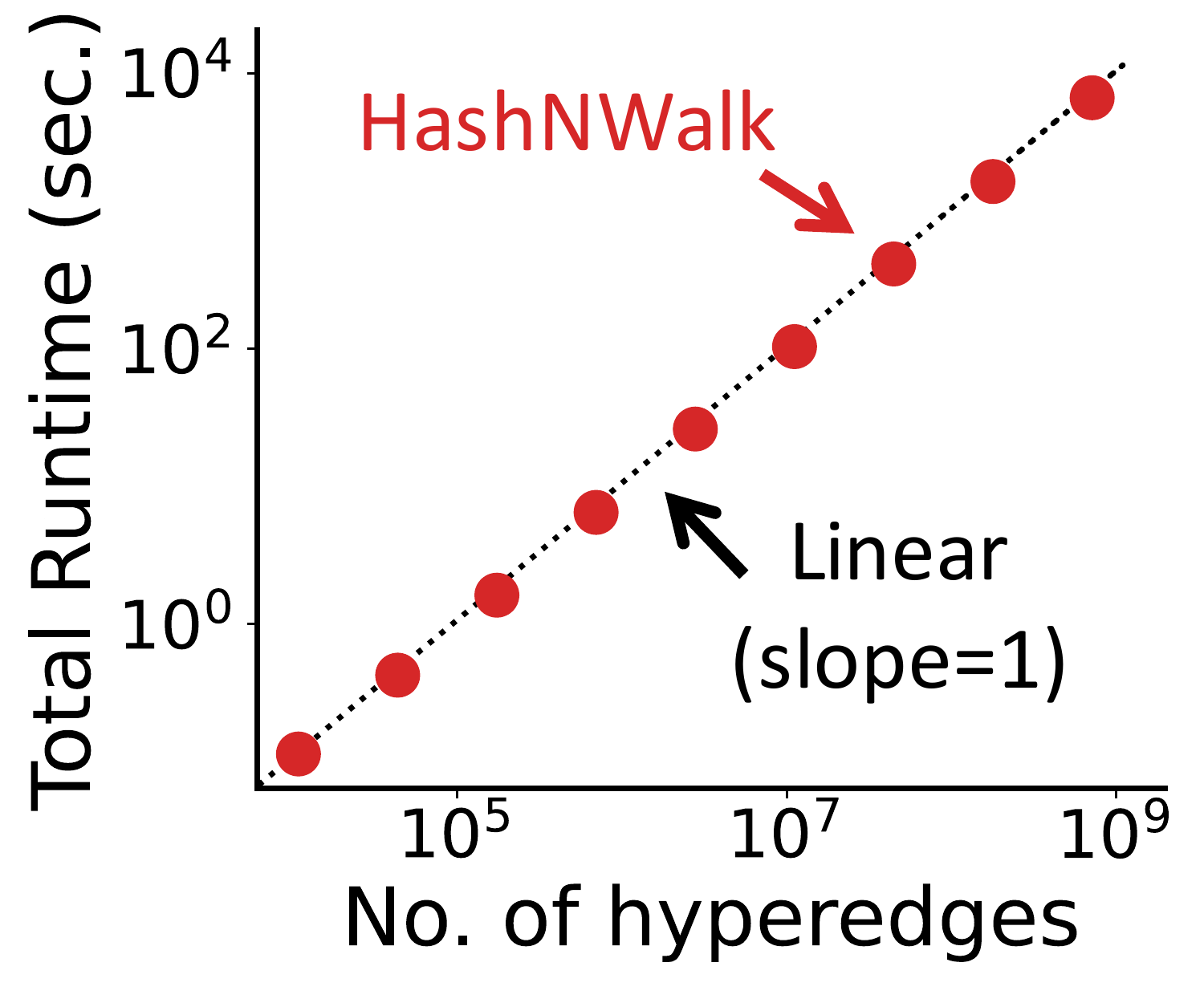}
		  \vspace{-5pt}
          \caption{Scalability}
          \label{fig:crown:scalability}
    \end{subfigure}
    \hspace{0pt}
    \begin{subfigure}[b]{.445\textwidth}
          \centering
          \includegraphics[width=0.995\columnwidth]{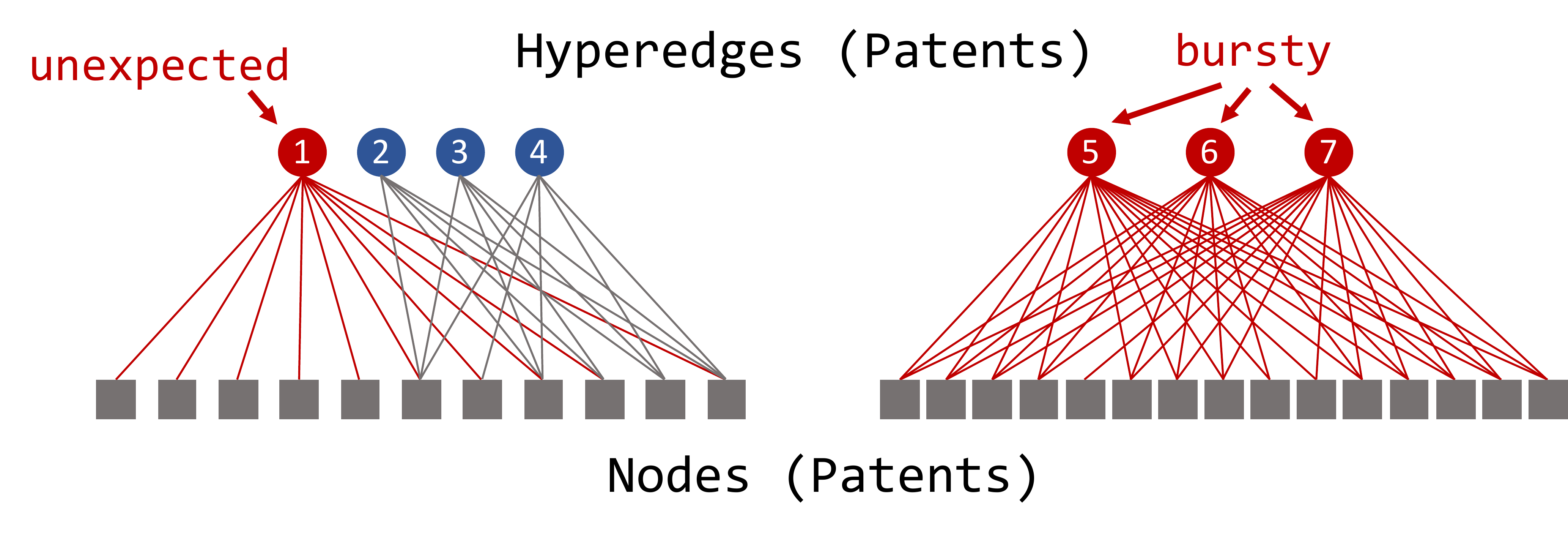}
		  \vspace{-15pt}
          \caption{Effectiveness}
          \label{fig:crown:effectiveness}
    \end{subfigure}
    \vspace{-8pt}
	\caption{\label{fig:crown} \underline{\smash{Strengths of \method.}} 
	(a) \method spots anomalous hyperedges  rapidly and precisely in a real-world hypergraph. (b) The total runtime of \method is linear in the size of the input hyperedge stream. (c) \method detects interesting patents. Patent 1 cited multiple patents that have not been cited together before, and patents 5-7 cited almost the same set of patents. See Section~\ref{sec:experiments} for details.}
	\vspace{-14pt}
\end{figure*}

A variety of real-world graphs, including computer networks, online social networks, and hyperlink networks, have been targets of attacks. 
Distributed denial-of-service attacks block the availability by causing an unexpected traffic jam on the target machine. In addition, fake connections in online social networks degrade the quality of recommendations, and those in hyperlink networks manipulate the centrality of webpages.
Due to its importance and necessity in real-world applications, anomaly detection in graphs has received considerable attention. To detect nodes, edges, and/or subgraphs deviating from structural and temporal patterns in graphs, various numerical measures of the deviation have been proposed with search algorithms \cite{akoglu2010oddball,hooi2016fraudar,shin2016corescope}. 
As many real-world graphs evolve over time,  detecting anomalies in real-time, as they appear, is desirable~\cite{bhatia2020midas,eswaran2018sedanspot}. 

While graphs model pairwise interactions,
interactions in many real-world systems are groupwise (collaborations of co-authors, group interactions on online Q\&A platforms, co-purchases of items, etc).
Such a groupwise interaction is naturally represented as a \textit{hyperedge}, i.e., a set of an arbitrary number of nodes.
A \textit{hypergraph}, which is a set of hyperedges, is an indispensable extension of a graph, which can only describe pairwise relations.
Moreover, many of such real-world hypergraphs evolve over time (e.g., emails exchanged continuously between sets of users, co-authorships established over time, and daily records of co-purchased items), and thus they are typically modeled as a stream of hyperedges. 

Despite the great interest in anomaly detection in graphs, the same problem in hypergraphs has been largely unexplored. High-order relationships represented by hyperedges exhibit structural and temporal properties distinguished from those in graphs and hence raise unique technical challenges. Thus, instead of simply decomposing hyperedges into pairwise edges and applying graph-based methods, it is required to take the underlying high-order structures into consideration for anomaly detection in hypergraphs. 


To this end, we propose \method, an online algorithm for detecting anomalous hyperedges.
\method maintains a constant-size summary that tracks structural and temporal patterns in high-order interactions in the input stream. Specifically, \method incorporates so-called \textit{edge-dependent node weights}~\cite{chitra2019random} into random walks on hypergraphs to estimate the proximity between nodes while capturing high-order information. Furthermore, we develop an incremental update scheme, which each hyperedge is processed by as it appears. 

The designed hypergraph summary is used to score the anomalousness of any new hyperedge in the stream. While the definition of anomaly depends on the context, in this work, we focus on two intuitive aspects: \textit{unexpectedness} and \textit{burstiness}. We assume that unexpected hyperedges consist of unnatural combinations of nodes, and bursty hyperedges suddenly appear in a short period of time. Based on the information in the form of a hypergraph summary, we formally define two anomaly score metrics that effectively capture these properties. 
We empirically show that \method is effective in detecting anomalous hyperedges in (semi-)real hypergraphs.

In summary, our contributions are as follows:

\begin{itemize}[leftmargin=*]
\item \textbf{Fast:} It takes a very short time for \method to process each new hyperedge. Specifically, in our experimental setting, it processed 1.4 billion hyperedges within 2.5 hours.
\item \textbf{Space Efficient:} The user can bound the size of the summary, which \method maintains. 
\item \textbf{Accurate:} \method successfully detects anomalous hyperedges. Numerically, it outperforms its state-of-the-art competitors with up to $47\%$ higher AUROC.
\end{itemize}

\noindent \textbf{Reproducibility:} The source code and datasets are available at \url{https://github.com/geonlee0325/HashNWalk}.


	
	\section{Related Works}
	\label{sec:related}
	We discuss prior works on the three topics relevant to this paper: (a) anomaly detection in (hyper)graphs; (b) summarization of edge streams; and (c) hypergraphs and applications.

\smallsection{Anomaly Detection in Graphs \& Hypergraphs:}
The problem of detecting anomalous nodes, edges, and/or subgraphs has been extensively studied for both static and dynamic graphs~\cite{akoglu2015graph}. In static graphs, nodes whose ego-nets are structurally different from others~\cite{akoglu2010oddball}, edges whose removal significantly reduces the encoding cost~\cite{chakrabarti2004autopart}, or subgraphs whose density is abnormally high~\cite{beutel2013copycatch,hooi2016fraudar,shin2016corescope} are assumed to be anomalies. 
In dynamic graphs, temporal edges are assumed to be anomalous if they connect sparsely connected parts in graphs~\cite{eswaran2018sedanspot} or are unlikely to appear according to underlying models~\cite{aggarwal2011outlier,yoon2019fast,bhatia2020midas,belth2020mining}. In addition, dense subgraphs generated within a short time are considered to be anomalous~\cite{shin2017densealert,eswaran2018spotlight}. Recently, embedding based methods have shown to be effective in detecting anomalies in graphs~\cite{yu2018netwalk,chang2020f}.

On the other hand, detecting anomalies in hypergraphs is relatively unexplored. 
Anomalous nodes in the hypergraph have been the targets of detection by using scan statistics on hypergraphs~\cite{park2009anomaly} or training a classifier based on the high-order structural features of the nodes~\cite{leontjeva2012fraud}.
The anomalousness of unseen hyperedges is measured based on how likely the combinations of nodes are drawn from the distribution of anomalous co-occurrences, which is assumed to be uniform, instead of the distribution of nominal ones~\cite{silva2008hypergraph}.
Approximate frequencies of structurally similar hyperedges obtained by locality sensitive hashing are used to score the anomalousness of hyperedges in the hyperedge stream~\cite{ranshous2017efficient}.
In this paper, we compare ours with the methods that detect anomalous interactions in online settings, 
i.e., anomaly detectors designed for edge streams~\cite{bhatia2020midas,eswaran2018sedanspot,chang2020f} and hyperedge streams~\cite{ranshous2017efficient}.


\smallsection{Summarization of Edge Streams:} Summarization aims to reduce the size of a given graph while approximately maintaining its structural properties. 
It has been particularly demanded in the context of real-time processing of streaming edges. 
In \cite{bhatia2020midas}, a count-min-sketch is maintained for approximate frequencies of edges.
Edge frequencies have been used to answer queries regarding structural properties of graphs~\cite{zhao2011gsketch,tang2016graph}. 
In \cite{bandyopadhyay2016topological}, local properties, such as the number of triangles and neighborhood overlap, are estimated by maintaining topological information of a given graph.

\smallsection{Hypergraphs and Applications:} 
Hypergraphs appear in numerous fields, including bioinformatics~\cite{hwang2008learning}, circuit design~\cite{karypis1999multilevel}, computer vision \cite{huang2009video,kim2020hypergraph}, natural language processing \cite{ding2020more}, social network analysis~\cite{yang2019revisiting}, and recommendation~\cite{mao2019multiobjective}.
Structural properties \cite{benson2018simplicial,do2020structural,lee2020hypergraph,lee2021how,choe2022midas} and dynamical properties \cite{benson2018simplicial,benson2018sequences,kook2020evolution,lee2021thyme+,choo2022persistence} of such real-world hypergraphs have been studied extensively.


	
	\vspace{-6pt}
	\section{Preliminaries}
	\label{sec:prelim}
	In this section, we introduce notations and preliminaries.


\subsection{Notations and Concepts}\label{sec:prelim:notations}

\smallsection{Hypergraphs:}
A \textit{hypergraph} $G=(V,E)$ consists of a set of nodes $V=\{v_1,...,v_{|V|}\}$ and a set of hyperedges $E=\{e_1,...,e_{|E|}\}$. 
Each hyperedge $e\in E$ is a non-empty subset of an arbitrary number of nodes. 
We can represent $G$ by its \textit{incidence matrix} $H\in \{0,1\}^{|E| \times |V|}$, where each entry $H_{ij}$ is $1$ if $v_j\in e_i$ and $0$ otherwise.
A \textit{hyperedge stream} $\{(e_i,t_i)\}_{i=0}^{\infty}$ is a sequence of hyperedges where each hyperedge $e_i$ arrives at time $t_i$.
For any $i$ and $j$, if $i<j$, then $t_i\leq t_j$.

\smallsection{Clique Expansion and Information Loss:}
\textit{Clique expansion}~\cite{zhou2007learning}, where each hyperedge $e\in E$ is converted to a clique composed of the nodes in $e$, is one of the most common ways of transforming a hypergraph $G$ into an ordinary pairwise graph. 
Clique expansion suffers from the loss of information on high-order interactions. That is, in general, a hypergraph is not uniquely identifiable from its clique expansion. Exponentially many non-isomorphic hypergraphs are reduced to identical clique expansions.

\smallsection{Random Walks on Hypergraphs:}
A random walk on a hypergraph $G$ is formulated in \cite{chitra2019random} as follows. If the current node is $u$, \textbf{(1)} select a hyperedge $e$ that contains the node $u$ (i.e., $u\in e$) with probability proportional to $\omega(e)$ and \textbf{(2)} select a node $v\in e$ with probability proportional to $\gamma_{e}(v)$ and walk to node $v$.
The weight $\omega(e)$ is the weight of the hyperedge $e$, and the weight $\gamma_e(v)$ is the weight of node $v$ with respect to the hyperedge $e$. 
The weight $\gamma_e(v)$ is \textit{edge-independent} if it is identical for every hyperedge $e$; and otherwise, it is \textit{edge-dependent}.
If all node weights are edge-independent, then a random walk on $G$ becomes equivalent to a random walk on its clique expansion~\cite{chitra2019random}. 
However, if node weights are edge-dependent, random walks on hypergraphs are generally \textit{irreversible}. That is, they may not be the same as random walks on any undirected graphs. 
In this sense, if edge-dependent weights are available, random walks are capable of exploiting high-order information beyond clique expansions and thus empirically useful in many machine learning tasks \cite{hayashi2020hypergraph}.


\smallsection{Transition Matrix:} 
To incorporate edge-dependent node weights, the incidence matrix $H$ is generalized to a weighted incidence matrix $R\in \mathbb{R}_{\geq 0}^{|E|\times |V|}$ where each entry $R_{ij}$ is $\gamma_{e_i}(v_j)$ if $v_j\in e_i$ and $0$ otherwise.
Then, the transition probability matrix $P\in \mathbb{R}^{|V|\times|V|}$ of a random walk on the hypergraph $G$ is written as $P = D_{V}^{-1} W D_{E}^{-1} R$, where $W\in \mathbb{R}^{|V|\times|E|}$ denotes the hyperedge-weight matrix where each entry $W_{ji}$ is $\omega(e_i)$ if $v_j\in e_i$ and 0 otherwise. The matrices $D_V\in \mathbb{R}^{|V|\times|V|}$ and $D_E\in \mathbb{R}^{|E|\times|E|}$ are diagonal matrices of node degrees and hyperedge weights, respectively. That is, 
if we let $q\in \mathbb{R}^{|E|}$ and $r\in \mathbb{R}^{|V|}$ be the vectors whose entries are all ones, then
$D_{V}=\text{diag}(Wq)$ and $D_{E}=\text{diag}(Rr)$. 

\subsection{Problem Description}\label{sec:prelim:problem}
The problem that we address in this paper is as follows.

\begin{pro}\label{problem}
    Given a stream $\{(e_i,t_i)\}_{i=1}^{\infty}$ of hyperedges, detect anomalous hyperedges, whose structural or temporal properties deviate from general patterns, in \textbf{near real-time} using \textbf{constant space}.
\end{pro}

\noindent While the definition of anomalous hyperedges depends on the context, we focus on two intuitive perspectives. 
In one aspect, a hyperedge is anomalous if it consists of an \textit{unexpected} subset of nodes. That is, we aim to detect hyperedges composed of unusual combinations of nodes. 
In the other aspect, we aim to identify a set of similar hyperedges that appear \textit{in bursts} as an anomaly. The sudden emergence of similar interactions often indicates malicious behavior harmful in many applications.
In addition, for time-critical applications, we aim to detect such anomalous hyperedges in near real-time, as they appear, using bounded space. 
While one might tempt to reduce hyperedges into subgraphs and solve the problem as anomalous subgraph detection, this harms the high-order information of the hyperedges. Also, existing works on anomalous subgraphs assume static graphs~\cite{hooi2016fraudar} or detect only the single most anomalous subgraph~\cite{shin2017densealert}, while we aim to score every hyperedge in the stream.

    \section{Proposed Method}
    \label{sec:method}
    \begin{figure*}[t]
	\vspace{-4mm}
	\centering
	\includegraphics[width=1.01\linewidth]{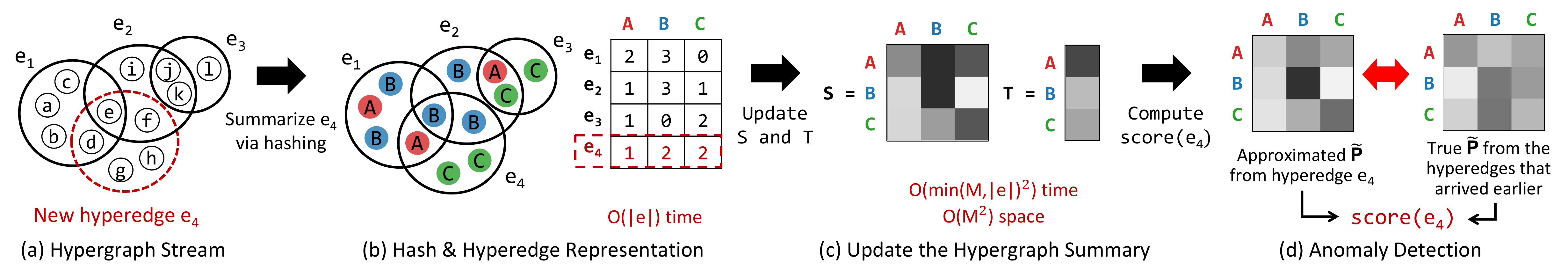}
	\vspace{-17pt}
	\caption{\label{fig:hashnwalk} \underline{\smash{Outline of \method.}} 
	(a) A new hyperedge arrives in the input hyperedge stream. (b) Nodes are merged into $M$ supernodes with edge-dependent weights by hashing, and hyperedges, including the new one, are represented as $M$-dimensional vectors ($M$=3 in this example). (c) The hypergraph summary is composed of a matrix $S$ and a vector $T$, and it is incrementally updated in response to the new hyperedge. (d) Based on the summary $\tilde{P}$, which is immediately obtainable from $S$ and $T$ (Eq.~\eqref{eq:p_ij}), the anomalousness of the new hyperedge is measured using the proposed scoring functions (Eq.~\eqref{eq:anomaly}).}
	\vspace{-12pt}
\end{figure*}

In this section, we propose \method (Algorithm~\ref{alg:method}), which is a fast and space-efficient algorithm for detecting anomalies in a hyperedge stream. Our main focus is speed and space efficiency since \method is expected to process a potentially infinite stream. 
As illustrated in Figure~\ref{fig:hashnwalk}, it maintains a concise and informative summary of a hyperedge stream (Sect.~\ref{sec:method:sketching}), which is incrementally updated as each new hyperedge arrives (Sect.~\ref{sec:method:incremental}). Once the summary is updated, anomalous hyperedges are identified immediately based on two principled metrics (Sect.~\ref{sec:method:anomaly}). While \method is based on multiple summaries (Sect.~\ref{sec:method:multiple}), we assume that it consists of a single summary for ease of explanation.
\subsection{Hypergraph Summarization}\label{sec:method:sketching}



\smallsection{Hyperedge Representation:}
We describe how to concisely represent each hyperedge using constant space. Hyperedges, by definition, are flexible in their sizes, and it is non-trivial to represent each hyperedge using the same amount of space. To this end, we map each node into one of $M$ different values using a hash function $h(\cdot): V \rightarrow \{1,...,M\}$. 
We consider each hash value as a \textit{supernode} that contains the nodes with the same hash value. Due to hash collisions, a hyperedge may contain a supernode multiple times,
and the number of occurrences becomes 
the weight of the supernode with respect to the hyperedge.
Formally, we represent each hyperedge $e$ of \textit{any size} into a $M$-dimensional vector $m(e)\in \mathbb{Z}^{M}$, whose $k$\textsuperscript{th} element indicates the number of the nodes that are contained in $e$ and mapped into the hash value $k$ (i.e., $m_{k}(e) := \sum_{v\in e} \mathds{1}(h(v)=k)$).
It is also interpreted as the weight of the supernode $k$ with respect to the hyperedge $e$. We denote $\tilde{e}$ as the set of supernodes that hyperedge $e$ contains, i.e., $\tilde{e} := \{k\, |\, m_k(e) > 0\}$.
Note that a hyperedge of any size is represented as a fixed-size vector, whose size $M$ is user-controlled.
In addition, the edge-dependent weights of supernodes can be utilized by random walks (see Section~\ref{sec:prelim:notations}).
If we use a constant-time hash function $h$ and a sparse vector format, for each hyperedge $e$, the time complexity of generating the vector $m(e)$ is $O(|e|)$, as stated in Lemma~\ref{lma:gen_m}. 

\begin{lma}[Time Complexity of Generating $m(e)$]~\label{lma:gen_m}
    Given a hyperedge $e$, it takes $O(|e|)$ time to generate the vector $m(e)$.
    
    \noindent\textsc{\textbf{Proof.}} Creating a zero vector in a sparse format (e.g., a hash table) and incrementing $m_{h(v)}(e)$ for every node $v\in e$ takes $O(|e|)$ time. \hfill \qedsymbol
\end{lma}

\begin{algorithm}[t]
\small
    \caption{\method}
    \label{alg:method}
    \begin{algorithmic}[1]
        \REQUIRE{\textbf{(1)} hyperedge stream: $\mathcal{E}=\{(e_i,t_i)\}_{i=1}^{\infty}$, \textbf{(2)} number of supernodes $M$, \textbf{(3)} number of hash functions $K$, \textbf{(4)} time-decaying parameter $\alpha$}
        \ENSURE{stream of anomaly scores $\{y_i\}_{i=1}^{\infty}$}
        \STATE{$S\in \mathbb{R}^{M\times M}$ and $T\in \mathbb{R}^{M}$\hfill \blue{$\triangleright$ Initialize to zeros}}
        \FOR{\upshape\textbf{each} hyperedge $(e_i,t_i)\in\mathcal{E}$}
            \STATE{$m(e_i)\leftarrow$ summarize $e_i$ via hashing \hfill\blue{$\triangleright$ Sect.~\ref{sec:method:sketching}}}
            \STATE{update $S$ and $T$ \hfill\blue{$\triangleright$ Sect.~\ref{sec:method:incremental}}}
            \STATE{$y_i\leftarrow \left(\mathsf{score_U}(e_i),\mathsf{score_B}(e_i)\right)$\hspace{293pt}\blue{$\triangleright$ Sect.~\ref{sec:method:anomaly}} }
        \ENDFOR
        \RETURN $\{y_i\}_{i=1}^{\infty}$
    \end{algorithmic}
\end{algorithm}

\smallsection{Hypergraph Summary:}
Below, we describe how to summarize the entire hypergraph for rapid and accurate anomaly detection.
We note that the key building block for identifying anomalous hyperedges of both types (i.e., unexpected ones and similar ones in bursts) is to estimate the proximity or structural similarity between nodes.
Thus, we summarize the input hypergraph in the form of proximity between supernodes, and we extend random walks to measure the proximity.
Our summary is based on random walks extended with edge-dependent supernode weights and hyperedge weights; and we use the transition probabilities as their approximation for rapid updates (see Section~\ref{sec:method:incremental}).
Specifically, we summarize the input hypergraph as a matrix $\tilde{P} := \tilde{D}_{V}^{-1} \tilde{W} \tilde{D}_{E}^{-1} \tilde{R}\in \mathbb{R}^{M\times M}$, where $\tilde{R}\in \mathbb{R}^{|E|\times M}$ is the weighted incidence matrix where each entry $\tilde{R}_{i\tilde{v}}$ is $\gamma_{\tilde{e}}(\tilde{v})$ if $\tilde{v}\in \tilde{e}_i$ and 0 otherwise.
The matrix $\tilde{W}\in \mathbb{R}^{M\times |E|}$ denotes the hyperedge-weight matrix where $\tilde{W}_{\tilde{v}\tilde{e}}$ is $\omega(\tilde{e})$ if $\tilde{v}\in \tilde{e}$ and 0 otherwise. 
The matrices $\tilde{D}_{V}\in \mathbb{R}^{M\times M }$ and $\tilde{D}_{E}\in \mathbb{R}^{|E|\times|E|}$ are diagonal matrices of supernode degrees and hyperedge weights, respectively.
Then, $\tilde{P}$ is the transition probability matrix where each entry $\tilde{P}_{\tilde{u}\tilde{v}}$ is the transition probability from supernode $\tilde{u}$ to $\tilde{v}$:

\begin{equation}~\label{eq:p}
    \tilde{P}_{\tilde{u}\tilde{v}} = \sum\nolimits_{i=1}^{|E|} \frac{\omega(\tilde{e}_i)\cdot\mathds{1}(\tilde{u}\in \tilde{e}_i)}{\tilde{W}_{\tilde{u}}} \cdot \frac{\gamma_{\tilde{e}_i}(\tilde{v})}{\tilde{R}_{\tilde{e}_i}}
\end{equation}
 
\noindent where $\tilde{W}_{\tilde{u}}$ is the weighted degree of the $\tilde{u}$, i.e., $\tilde{W}_{\tilde{u}} := \sum_{i=1}^{|E|}\omega(\tilde{e}_i)\cdot\mathds{1}(\tilde{u}\in \tilde{e}_i)$, and $\tilde{R}_{\tilde{e}_i}$ is the sum of the weights of the supernodes in the $\tilde{e}_i$, i.e., $\tilde{R}_{\tilde{e}_i} := \sum_{\tilde{v}\in \tilde{e}_i}\gamma_{\tilde{e}_i}(\tilde{v})$. 
 
\smallsection{Edge-Dependent Supernode Weights:} 
If edge-dependent supernode weights are available, random walks utilize high-order information beyond clique expansions. Such weights are naturally obtained from the aforementioned vector representation of hyperedges. 
That is, we use the number of the occurrences of each supernode $\tilde{v}$ in each hyperedge $\tilde{e_i}$ as the weight of $\tilde{v}$ with respect to $\tilde{e_i}$. Formally, $\gamma_{\tilde{e}_i}(\tilde{v})=m_{\tilde{v}}(e_i)$, and thus $R_{\tilde{e}_i} = \sum_{\tilde{v}\in \tilde{e}_i}\gamma_{\tilde{e}_i}(\tilde{v}) = \sum_{k=1}^{M} m_k(e_i)=|e_i|$.

 

\smallsection{Time-Decaying Hyperedge Weights:}
In order to facilitate identifying recent bursts of similar hyperedges, which are one of our focuses, we emphasize recent hyperedges with large weights.
Specifically, at current time $t$, we define the weight of each hyperedge $e_i$, which is arrived at time $t_i$, as $\omega(e_i)=\ker(t-t_i)=\alpha^{t-t_i}(1-\alpha)$ where $\ker(x):=\alpha^x(1-\alpha)$ is a kernel function for quantifying time decay and $\alpha\in [0,1)$ is a hyperparameter that determines the degree of emphasis. 
Specifically, smaller $\alpha$ more emphasizes recent hyperedges. 
\subsection{Incremental Update}\label{sec:method:incremental}
\smallsection{Challenges:}
Constructing $\tilde{P}$ from scratch, which takes $O(|E|\cdot M^2)$ time, is undesirable when immediate responses to anomalies are demanded.
In addition, when hyperedges are streamed indefinitely,  materializing $\tilde{W}$, $\tilde{D}_E$, and $\tilde{R}$, which are used to compute $\tilde{P}$, is prohibitive since their sizes are proportional to the number of hyperedges.

\smallsection{Proposed Updated Scheme:}
We present an incremental algorithm for efficiently but exactly updating $\tilde{P}$ in response to a new hyperedge. The proposed update scheme maintains only $\tilde{P}$, whose size is controllable by the user, without materializing any larger matrix. 
Assume $m$ hyperedges $e_1,...,e_m$ have arrived,  
 and let $\tilde{P}_{\tilde{u}\tilde{v}}^{(m)}$ be the proximity from supernode $\tilde{u}$ to supernode $\tilde{v}$ in them. 
We introduce a matrix $S\in \mathbb{R}^{M\times M}$ and a vector $T\in \mathbb{R}^{M}$, and for any supernodes $\tilde{u}$ and $\tilde{v}$, their entries when the hyperedge $e_m$ arrives at time $t_m$ are
\begin{align*}
    & S_{\tilde{u}\tilde{v}}^{(m)}:=\sum\nolimits_{i=1}^{m}\alpha^{-t_i}\cdot \mathds{1}(\tilde{u}\in \tilde{e}_i)\cdot \frac{\gamma_{\tilde{e}_i}(\tilde{v})}{\tilde{R}_{\tilde{e}_i}}
\end{align*}
\begin{align*}
    & T_{\tilde{u}}^{(m)}:=\sum\nolimits_{i=1}^{m}\alpha^{-t_i}\cdot \mathds{1}(\tilde{u}\in \tilde{e}_i),
\end{align*}

\noindent
Then, based on Eq.~\eqref{eq:p} and the predefined hyperedge weight function $\ker(x)=\alpha^x(1-\alpha)$, $\tilde{P}_{\tilde{u}\tilde{v}}^{(m)}$ is written as

\vspace{-10pt}
\begin{equation}
    \tilde{P}_{\tilde{u}\tilde{v}}^{(m)} = \frac{\sum\limits_{i=1}^{m} \alpha^{t_m-t_i}(1-\alpha)\cdot \mathds{1}(\tilde{u}\in \tilde{e}_i)\cdot \frac{\gamma_{\tilde{e}_i}(\tilde{v})}{\tilde{R}_{\tilde{e}_i}}}{\sum\limits_{i=1}^{m}\alpha^{t_m-t_i}(1-\alpha)\cdot \mathds{1}(\tilde{u}\in \tilde{e}_i)}=\frac{S_{\tilde{u}\tilde{v}}^{(m)}}{T_{\tilde{u}}^{(m)}}.\label{eq:p_ij}
\end{equation}

\noindent Instead of directly tracking the proximity matrix $\tilde{P}$, we track  aforementioned $S$ and $T$, whose entries are initialized to zero.
Each entry $S_{\tilde{u}\tilde{v}}$ and $T_{\tilde{u}}$ can be updated in constant time, as presented in Lemmas~\ref{lma:upd_S} and \ref{lma:upd_T}, and once they are updated, we can compute $\tilde{P}_{\tilde{u}\tilde{v}}^{(m)}$ in $O(1)$ time by Eq.~\eqref{eq:p_ij}, if necessary.

\begin{lma}[Updating $S_{\tilde{u}\tilde{v}}$]~\label{lma:upd_S}
For any $m\geq 0$, when the hyperedge $e_{m+1}$ arrives at $t_{m+1}$, Eq.~\eqref{eq:upd_S} holds.
    \begin{equation}\label{eq:upd_S}
        S_{\tilde{u}\tilde{v}}^{(m+1)} = S_{\tilde{u}\tilde{v}}^{(m)} + \alpha^{-t_{m+1}}\cdot \mathds{1}(\tilde{u}\in \tilde{e}_{m+1})\cdot \frac{\gamma_{\tilde{e}_{m+1}}(\tilde{v})}{\tilde{R}_{\tilde{e}_{m+1}}}.
    \end{equation}
\end{lma}

\begin{lma}[Updating $T_{\tilde{u}}$]~\label{lma:upd_T}
    For any $m\geq 0$, when the hyperedge $e_{m+1}$ arrives at $t_{m+1}$, Eq.~\eqref{eq:upd_T} holds.
    \begin{equation}\label{eq:upd_T}
        T_{\tilde{u}}^{(m+1)} = T_{\tilde{u}}^{(m)} + \alpha^{-t_{m+1}} \cdot \mathds{1}(\tilde{u} \in \tilde{e}_{m+1}).
    \end{equation}
\end{lma}

\noindent Lemma~\ref{lma:upd_S} and Lemma~\ref{lma:upd_T} are immediate from the definitions of $S_{\tilde{u}\tilde{v}}^{(m)}$ and $T_{\tilde{u}}^{(m)}$.



\smallsection{Complexity:}
Notably, if $\tilde{u} \notin \tilde{e}_{m+1}$, $\mathds{1}(\tilde{u}\in \tilde{e}_{m+1})=0$ holds and if $\tilde{v} \notin \tilde{e}_{m+1}$, $\gamma_{\tilde{e}_{m+1}}(\tilde{v})=0$ holds. Thus, if $\tilde{u}$ or $\tilde{v}$ is not included in the new hyperedge (i.e., $\tilde{u}\notin \tilde{e}_{m+1}$ or $\tilde{v}\notin \tilde{e}_{m+1}$), $S_{\tilde{u}\tilde{v}}$ remains the same (i.e., $S_{\tilde{u}\tilde{v}}^{(m+1)}=S_{\tilde{u}\tilde{v}}^{(m)}$) and thus does not need any update. Similarly, $T_{\tilde{u}}$ does not change if $\tilde{u}$ is not included in $\tilde{e}_{m+1}$. These facts significantly reduce the update time of the summary, enabling near real-time processing of each hyperedge. To sum up, in response to a new hyperedge, \method updates the summary in a short time using constant space, as stated in Lemmas~\ref{lma:upd_time} and \ref{lma:memory}, respectively.

\begin{lma}[Update Time Per Hyperedge]~\label{lma:upd_time}
    Given the sparse vector representation $m(e)$ of a hyperedge $e$, updating $S\in \mathbb{R}^{M\times M}$ and $T\in \mathbb{R}^{M}$ using Eq.~\eqref{eq:upd_S} and Eq.~\eqref{eq:upd_T} takes $O(\min(M,|e|)^2)$ time.
    
\noindent\textsc{\textbf{Proof.}}
The number of supernodes in $e$ is $|\tilde{e}|$, which is at most the number of nodes $|e|$ and the number of supernodes $M$, and thus $|\tilde{e}| = O(\min(M,|e|))$. Then, $|\tilde{e}|^2$ elements of $S$ and $|\tilde{e}|$ elements of $T$ are updated by Eq.~\eqref{eq:upd_S} and Eq.~\eqref{eq:upd_T}, and the update time is constant per element. Therefore, the total time complexity is $O(\min(M,|e|)^2)$. \hfill $\blacksquare$
\end{lma}

\begin{lma}[Constant Space] \label{lma:memory}
    The maintained summary $\tilde{P}$ takes $O(M^2)$ space.
    
\noindent\textsc{\textbf{Proof.}}
The matrix $S$ and the vector $T$ require $O(M^2)$ and $O(M)$ space, respectively. 
\hfill $\blacksquare$
\end{lma}


\subsection{Anomaly Detection}\label{sec:method:anomaly}

\smallsection{Hyperedge Anomaly Score:}
We now propose an online anomalous hyperedge detector, which is based on the structural and temporal information captured in the summary $\tilde{P}$.
We evaluate each newly arriving hyperedge by measuring a hyperedge anomaly score defined in Definition~\ref{defn:anomaly}. 

\begin{dfn}[Hyperedge Anomaly Score]
    Given a newly arriving hyperedge $e_i$ at time $t_i$, its anomaly score is defined as \label{defn:anomaly}
    \begin{equation}
        \mathsf{score}(e_i) = \underset{\tilde{u},\tilde{v}\in \tilde{e}_i}{\mathsf{aggregate}} \left( d_{\tilde{u},t_i}^\beta \cdot \log \frac{a_{\tilde{u}\tilde{v}}}{s_{\tilde{u}\tilde{v}}}\right), \label{eq:anomaly}
    \end{equation}
    where $d_{\tilde{u},t_i}$ is the number of occurrences of $\tilde{u}$ at time $t_i$, $\beta\in[0,\infty)$ is a hyperparameter for the importance of the occurrences,
    $a_{\tilde{u},\tilde{v}}=\frac{\gamma_{\tilde{e}_i}(\tilde{v})}{\tilde{R}_{\tilde{e}_i}}$, and $s_{\tilde{u},\tilde{v}}$ is $\tilde{P}_{\tilde{u},\tilde{v}}$ just before $t_i$.
    Intuitively, $a_{\tilde{u},\tilde{v}}$ and $s_{\tilde{u},\tilde{v}}$ are  the ``observed'' proximity (i.e., the proximity in the current hyperedge)  
    and ``expected'' proximity (i.e., the proximity in all past hyperedges appearing before $t_i$) from supernode $\tilde{u}$ to supernode $\tilde{v}$, respectively.
\end{dfn}
\noindent
Note that the relationships between all pairs of supernodes in the hyperedge, including the pairs of the same supernode, are taken into consideration, and they are aggregated using any aggregation functions.
%
The hyperparameter $\beta$ and the $\textsf{aggregate}$ function can be controlled to capture various types of anomalies. For the two types of anomalies, we define scoring functions $\mathsf{score_U}$ and $\mathsf{score_B}$ as described below. 

\smallsection{Unexpectedness ($\mathsf{score_U}$):}
Intuitively, $a_{\tilde{u},\tilde{v}} / s_{\tilde{u},\tilde{v}}$ in Eq.~\eqref{eq:anomaly} measures how much the proximity from the supernode $\tilde{u}$ to $\tilde{v}$ in the new hyperedge $e_i$ deviates from the proximity in the past hyperedges. Specifically, the ratio is high if two supernodes $\tilde{u}$ and $\tilde{v}$ that have been far from each other in past hyperedges unexpectedly co-appear with high proximity in the new hyperedge. 
Thus, in $\mathsf{score_U}$, which is the anomaly score for identifying unexpected hyperedges, we focus on the ratio by setting $\beta=0$.
In order to detect any such unexpected pairs of supernodes in the hyperedge, $\mathsf{score_U}$ uses the maximum ratio as the final score (i.e., $\textsf{aggregate}=\textsf{max}$).

\smallsection{Burstiness ($\mathsf{score_B}$):}
In order to detect similar hyperedges that appear in bursts, the number of occurrences of supernodes is taken into consideration. Supernodes, by definition, are subsets of nodes, and similar hyperedges tend to share many supernodes. If a large number of similar hyperedges appear in a short period of time, then the occurrences of the supernodes in them tend to increase accordingly. Thus, in $\mathsf{score_B}$, which is the anomaly score for identifying recent bursts of similar hyperedges, we set $\beta$ to a positive number (specifically, $1$ in this work) to take such occurrences (i.e., $d_{\tilde{u},t_i}^{\beta}$ in Eq.~\eqref{eq:anomaly}) into consideration, in addition to unexpectedness (i.e., $a_{\tilde{u},\tilde{v}} / s_{\tilde{u},\tilde{v}}$ in Eq.~\eqref{eq:anomaly}).
We reflect the degrees of all supernodes in the hyperedge by averaging the scores from all supernode pairs (i.e., $\textsf{aggregate}=\textsf{mean}$).

\smallsection{Complementarity of the Anomaly Scores:} 
While the only differences between $\mathsf{score_U}$ and $\mathsf{score_B}$ are the consideration of the current degree of supernodes (i.e., $d_{\tilde{u},t_i}^{\beta}$) and the aggregation methods, the differences play an important role in identifying specific types of anomalies (see Section~\ref{sec:experiment:accuracy}).


\smallsection{Complexity:}
For each new hyperedge $e$, \method computes $\mathsf{score}(e)$ in a short time, as stated in Lemma~\ref{lma:score}.

\begin{lma}[Scoring Time Per Hyperedge] \label{lma:score}
    Given the hypergraph summary $\tilde{P}$ and a hyperedge $e$ in the form of a vector $m(e)$, computing $\mathsf{score}(e)$ takes $O(\min(M,|e|)^2)$ time.

\noindent\textsc{\textbf{Proof.}}
The number of supernodes in $e$ is $O(\min(M,|e|))$.
We maintain and update the current degrees of supernodes, which takes $O(\min(M,|e|))$ time for each new hyperedge $e$.
There are $O(\min(M,|e|)^2)$ pairs of supernodes in $\tilde{e}$, and the computation for each supernode pair in Eq.~\eqref{eq:anomaly} takes $O(1)$ time. Hence, the total time complexity is $O(\min(M,|e|)^2)$.\hfill $\blacksquare$ \vspace{-1mm}
\end{lma}





\begin{thm}[Total Time Per Hyperedge]~\label{thm:total_time}
    \method takes $O(|e|+\min(M,|e|)^2)$ time to process a hyperedge $e$.
    \noindent\textsc{\textbf{Proof.}}
    Theorem~\ref{thm:total_time} follows from Lemmas~\ref{lma:gen_m}, \ref{lma:upd_time}, and \ref{lma:score}. \hfill $\blacksquare$
\end{thm}

\subsection{Using Multiple Summaries (Optional)}\label{sec:method:multiple}
Multiple hash functions can be used in \method to improve its accuracy at the expense of speed and space.
Specifically, if we use $K$ hash functions, maintain $K$ summaries, and compute $K$ scores independently, then the space and time complexities become $K$ times of those with one hash function. 
Given hyperedge anomaly scores from $K$ different summaries, we use the \textbf{maximum} one as the final score, although any other aggregation function can be used instead.
\vspace{-5pt}

    \section{Experiments}
    \label{sec:experiments}
    We review our experiments to answer Q1-Q4:
\begin{enumerate}[label=Q\arabic*.,leftmargin=*]
\item \textbf{Performance:} How rapidly and accurately does \method detect anomalous hyperedges?
\item \textbf{Discovery:} What meaningful events can \method detect in real-world hypergraph streams?
\item \textbf{Scalability:} How does the total runtime of \method change with respect to the input stream size?
\item \textbf{Parameter Analysis:} How do the parameters of \method affect its performance? 
\vspace{-5pt}
\end{enumerate}

\subsection{Experimental Settings}
\label{sec:experiments:settings}

\begin{table}[t]
    \small
	\begin{center}
		\caption{\label{tab:datasets} Five real-world hypergraphs. 
		}
		\vspace{-7pt}
		\scalebox{1}{
			\begin{tabular}{l|c|c|c|c}
				\toprule
				\textbf{Dataset} & $\mathbf{|V|}$ & $\mathbf{|E|}$ & $\mathbf{\mathrm{avg}_{e\in E}|e|}$ & $\mathbf{\max_{e\in E}|e|}$\\
				\midrule
				\texttt{Email-Enron} & 143 & 10,885 & 2.472 & 37 \\
				\texttt{Transaction} & 284,807 & 284,807 & 5.99 & 6 \\ 
				\texttt{DBLP} & 1,930,378 & 3,700,681 & 2.790 & 280\\
				\texttt{Cite-patent} & 4,641,021 & 1,696,554 & 18.103 & 2,076\\
				\texttt{Tags-overflow} & 49,998 & 14,458,875 & 2.968 & 5\\
				\bottomrule 
			\end{tabular}}
	\end{center}
	\vspace{-15pt}
\end{table}
    
\smallsection{Datasets:} We used five different real-world datasets in Table~\ref{tab:datasets}. They are described in detail in later subsections. 

\smallsection{Machines:}
We ran \ffade on a workstation with an Intel Xeon 4210 CPU, 256GB RAM, and RTX2080Ti GPUs. We ran the others on a desktop with an Intel Core i9-10900KF CPU and 64GB RAM.

\begin{figure*}[t]
	\vspace{-10pt}
	\centering
	\includegraphics[width=0.7\linewidth]{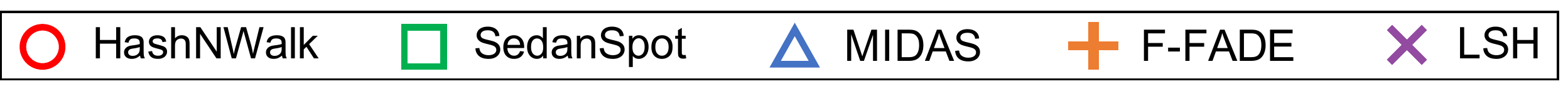}\\
    \begin{subfigure}[b]{.45\textwidth}
          \centering
          \includegraphics[width=0.985\columnwidth]{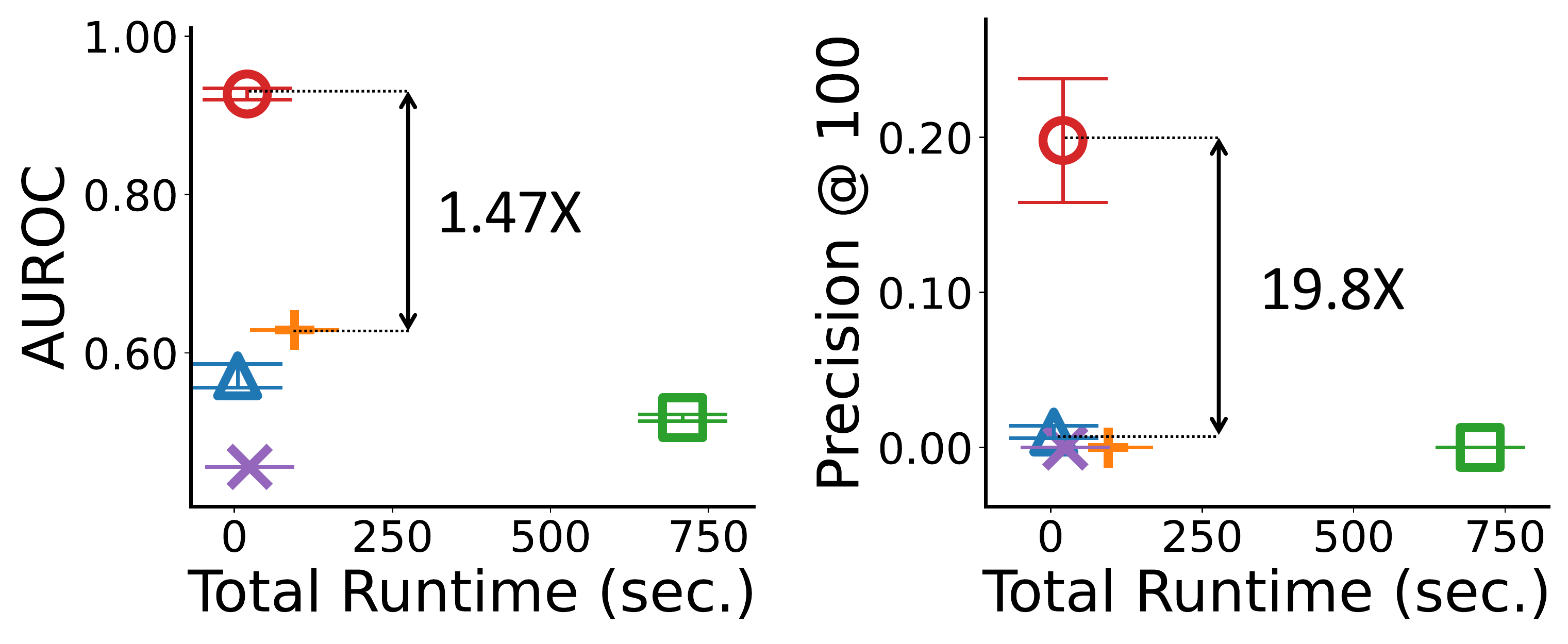}
		  \vspace{-5pt}
          \caption{\real}
          \label{fig:time:real}
    \end{subfigure}
    \hspace{0pt}
    \begin{subfigure}[b]{.45\textwidth}
          \centering
          \includegraphics[width=0.985\columnwidth]{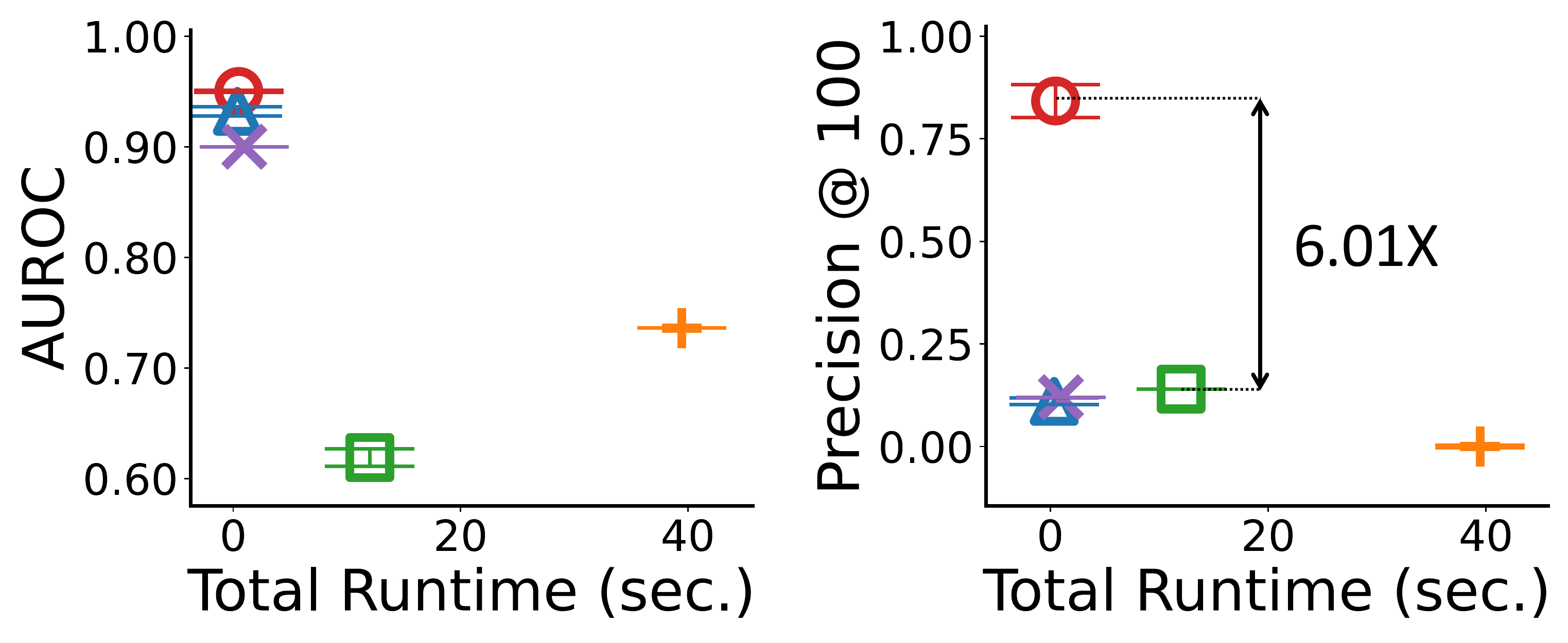}
		  \vspace{-5pt}
          \caption{\syntheticU}
          \label{fig:time:syntheticU}
    \end{subfigure}
    \hspace{0pt}
    \begin{subfigure}[b]{.45\textwidth}
          \centering
          \includegraphics[width=0.985\columnwidth]{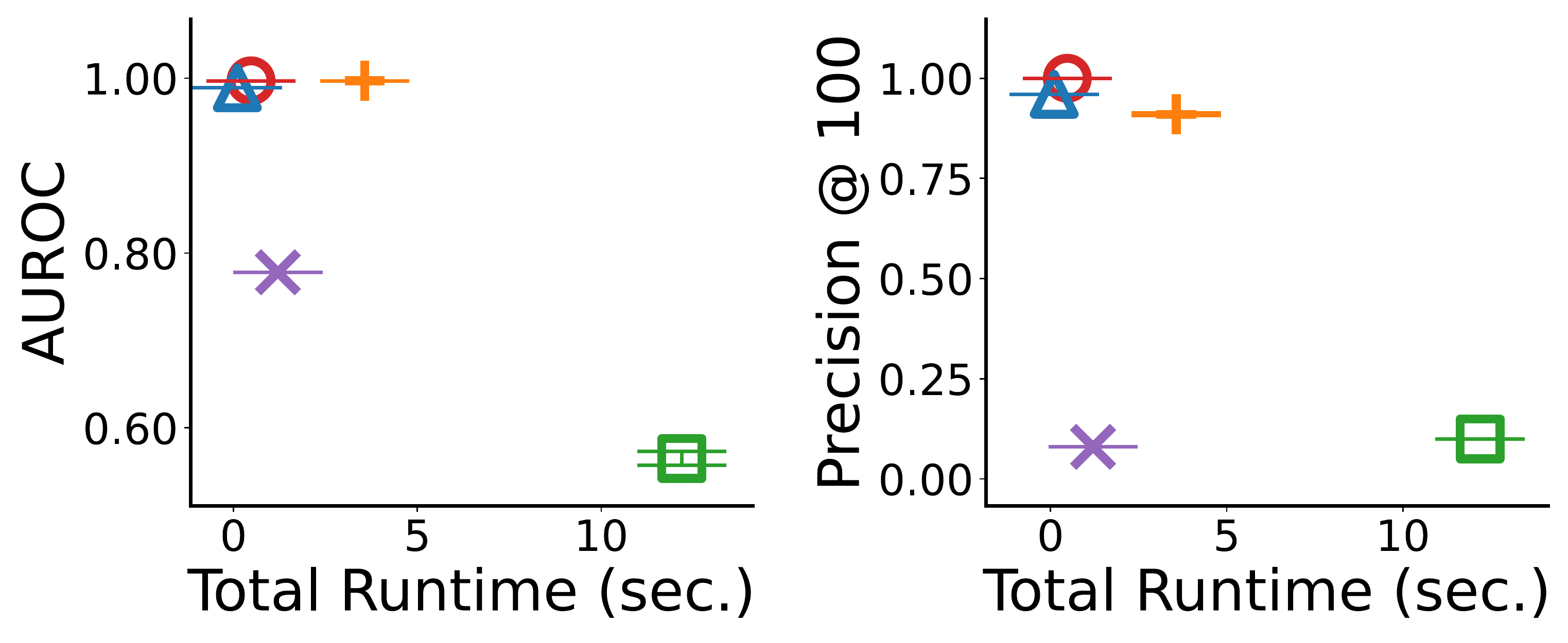}
		  \vspace{-5pt}
          \caption{\syntheticB}
          \label{fig:time:syntheticB}
    \end{subfigure}
    \vspace{-7pt}
	\caption{\label{fig:time} \method is accurate (in terms of AUROC and Prec.@100) and fast. For example, in the \real dataset, \method achieves $47\%$ higher AUROC with $4.7\times$ faster speed, compared to \ffade.} 
	\vspace{-8pt}
\end{figure*}

\smallsection{Baselines:}
We consider four streaming algorithms for anomaly detection in graphs and hypergraphs as competitors:
\begin{itemize}[leftmargin=*]
        \item \textbf{\sedanspot \cite{eswaran2018sedanspot}:} Given a stream of edges, it aims to detect \textit{unexpected edges}, i.e., edges that connect nodes from sparsely connected parts of the graph, based on personalized PageRank scores.
        \item \textbf{\midas \cite{bhatia2020midas}:} Given a stream of edges, it aims to detect \textit{similar edges in bursts}. To this end, it uses the Count-Min-Sketch. 
        \item \textbf{\ffade \cite{chang2020f}:} Given a stream of edges, it uses frequency-based matrix factorization and computes the likelihood-based anomaly score of each edge that combines \textit{unexpectedness} and \textit{burstiness}. 
        \item \textbf{\lsh \cite{ranshous2017efficient}:} Given a stream of hyperedges, it computes the \textit{unexpectedness} of each one using its approximate frequency so far. 
\end{itemize}
For graph-based anomaly detection methods, we transform hypergraphs into graphs via clique expansion (Section~\ref{sec:prelim:notations}). 
That is, each hyperedge $e_i$ is reduced to $|e_i|^2$ pairwise edges, and the timestamp $t_i$ is assigned to each edge. The anomaly score of the hyperedge is computed by aggregating the anomaly scores of the pairwise edges, using the best one among arithmetic/geometric mean, sum, and maximum.

\smallsection{Implementation:}
We implemented \method and \lsh in C++ and Python, respectively.
For the others, we used their open-source implementation.
\sedanspot and \midas are implemented in C++ and \ffade is implemented in Python. 



\smallsection{Evaluation:}
Given anomaly scores of hyperedges, we measure AUROC and Precision@$k$ (i.e., the ratio of true positives among $k$ hyperedges with the highest scores).


\subsection{Q1. Performance Comparison}\label{sec:experiment:accuracy}
We consider three hypergraphs: \real, \syntheticU, and \syntheticB. 
\real \cite{dal2015calibrating} is a real-world hypergraph of credit card transactions. 
Each timestamped transaction is described by a $28$ dimensional feature vector.
There exist $492$ frauds, which account for $0.172\%$ of the entire transactions. 
For each transaction, we generate a hyperedge by grouping it with $5$ nearest transactions that occurred previously. 
Thus, each node is a transaction, and each hyperedge is a set of transactions that are similar to each other. 

In \texttt{Email-Enron}, each node is an email account and each hyperedge is the set of the sender and receivers. The timestamp of each hyperedge is when the email was sent. 
We consider two scenarios \textsc{InjectionU} and \textsc{InjectionB}, where we generate two semi-real hypergraphs \syntheticU and \syntheticB by injecting 200 unexpected and bursty hyperedges, respectively, in \texttt{Email-Enron}. 
The two injection scenarios are designed as follows:


\noindent\fbox{%
        \parbox{0.98\columnwidth}{%
        \vspace{-2mm}
        \begin{itemize}[leftmargin=*]
    \item \textbf{\textsc{InjectionU}}: \underline{\smash{Injecting unexpected hyperedges.}}
    \begin{enumerate}[leftmargin=8pt]
    \itemsep0em 
        \item\small Select a hyperedge $(e_i,t_i)\in\mathcal{E}$ uniformly at random.
        \item\small Create a hyperedge by replacing $\lceil{|e_i|\,/\,2}\rceil$ nodes in $e_i$ with random ones, and set their timestamp to $t_i$.
        \item\small Repeat (1)-(2) $g$ times to generate $g$ hyperedges.
    \end{enumerate}
    \item \textbf{\textsc{InjectionB}}: \underline{\smash{Injecting bursty hyperedges.}}
    \begin{enumerate}[leftmargin=8pt]
        \item\small Select a time $t\in \{t_{\text{setup}}+1,\cdots, t_m\}$ uniformly at random.~\label{InjectionB:timestamp}
        \item\small Sample a set of $n$ nodes $N\subseteq V$ uniformly at random.~\label{InjectionB:sample}
        \item\small Create $m$ uniform random subsets of $N$ at time $t$. Their sizes are chosen uniformly at random from $\{1,\cdots,n\}$. 
        \item\small Repeat (1) - (3) $l$ times to generate $m\cdot l$ hyperedges. 
    \end{enumerate}
\end{itemize}
        \vspace{-2mm}
        }%
    }


\noindent
All anomalies are injected after time $t_{\text{setup}}$, and thus all methods are evaluated from time $t_{\text{setup}}$ where we set $t_{\text{setup}}=t_{100}$ ($<t_{|E|\cdot 0.01}$). In \textsc{InjectionU}, we set $g=200$. In \textsc{InjectionB}, we set $m=20$, $n=5$, and $l=10$.
    


\smallsection{Accuracy:}
In \real, we use $\mathsf{score_U}$ and set $\alpha=0.98$, $K=4$, and $M=350$. In \syntheticU and \syntheticB, we use $\mathsf{score_U}$ and $\mathsf{score_B}$, respectively, and commonly set $\alpha=0.98$, $K=15$, and $M=20$. 
As discussed later, these summaries take up less space than the original hypergraphs. 
As shown in Figure~\ref{fig:time}, \method accurately detects anomalous hyperedges in real and semi-real hypergraphs. 
Notably, while most methods fail to find any anomalous hyperedges in their top $50$ (see Figure~\ref{fig:crown:accuracy} in Section~\ref{sec:intro}) or top $100$ (Figure~\ref{fig:time:real}) hyperedges with the highest anomaly scores, \method is successful. 
In addition, \method accurately detects both unexpected and bursty hyperedges. Note that while several competitors successfully spot bursty hyperedges in \syntheticB, most of them fail to spot unexpected ones in \syntheticU. 
Specifically, in \syntheticU, \method achieves $6.01\times$ higher precision@100 with $27\times$ faster speed than \sedanspot.


\begin{table}[t!]
\small
\begin{center}
\caption{\label{tab:complementary} The two proposed hyperedge anomaly scoring metrics $\mathbf{\mathsf{score_U}}$ and $\mathbf{\mathsf{score_B}}$ complement each other.}
\vspace{-5pt}
    \scalebox{1}{
    \begin{tabular}{c||cc|cc}
        \toprule
        & \multicolumn{2}{c|}{\textbf{\syntheticU}} & \multicolumn{2}{c}{\textbf{\syntheticB}}\\
        & AUROC & Prec.@100 & AUROC & Prec.@100\\
        \midrule
        $\mathsf{score_U}$ & \textbf{0.951} & \textbf{0.815} & 0.802 & 0.740 \\
        $\mathsf{score_B}$ & 0.916 & 0.090 & \textbf{0.997} & \textbf{1.000} \\
    \bottomrule
    \end{tabular}}
\end{center}
\vspace{-10pt}
\end{table}

\smallsection{Speed:}
As seen in Figure~\ref{fig:time}, \method is one of the fastest methods among the considered ones. Notably, in \syntheticU, \method is $27\times$ faster than the second most accurate method.

\smallsection{Space Usage:}
We analyze the amount of space used by \method. 
Let $C_Z$ and $C_F$ be the numbers of bits to encode an integer and a floating number, respectively, and we assume $C_Z=C_F=32$.
The size of the original hypergraph $G=(V,E)$ is the sum of the hyperedge sizes, and precisely, $C_Z\cdot \sum_{e\in E}|e|$ bits are required to encode the hypergraph.
As described in Lemma~\ref{lma:memory} in Section~\ref{sec:method:incremental}, for each hash function, \method tracks a matrix $S\in \mathbb{R}^{M\times M}$ and a vector $T\in \mathbb{R}^{M}$, and thus it requires $C_F\cdot K\cdot(M^2+M)$ bits with $K$ hash functions. 
We set $K=4$ and $M=350$ in \real; and $K=15$ and $M=20$ in \syntheticU and \syntheticB.
As a result, \method requires about $28.6\%$ and $22.5\%$ of the space required for the original hypergraphs, in \real and semi-real hypergraphs, respectively.
For competitors, we conduct hyperparameter tuning, including configurations requiring more space than ours.\footnote{See \url{https://github.com/geonlee0325/HashNWalk} for details.}

\smallsection{Complementarity of $\mathsf{score_U}$ and $\mathsf{score_B}$:}
As seen in Table~\ref{tab:complementary}, 
while $\mathsf{score_U}$ is effective in detecting unexpected hyperedges, it shows relatively low accuracy in detecting bursty hyperedges.
The opposite holds in $\mathsf{score_B}$.
The results indicate that the two metrics $\mathsf{score_U}$ and $\mathsf{score_B}$ are complementary. 

\subsection{Q2. Discovery} \label{sec:experiment:discoveries}
Here, we share the results of case studies conducted on the \texttt{DBLP}, \texttt{Cite-patent}, and \texttt{Tags-overflow} datasets. 

\begin{figure}[t]
	\vspace{-2pt}
	\centering
    \begin{subfigure}[b]{.32\textwidth}
          \centering
          \includegraphics[width=0.9\columnwidth]{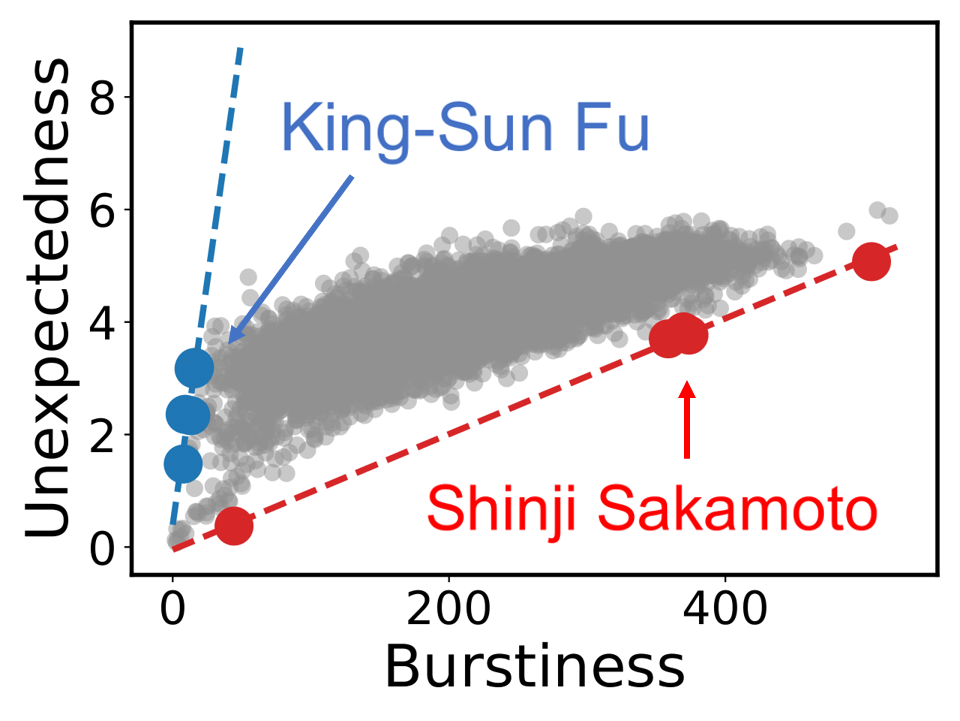}
		  \vspace{-5pt}
          \caption{$\mathsf{avg_U}$ and $\mathsf{avg_B}$}
          \label{fig:case:B}
    \end{subfigure}
    \hspace{5pt}
    \begin{subfigure}[b]{.32\textwidth}
          \centering
          \includegraphics[width=0.9\columnwidth]{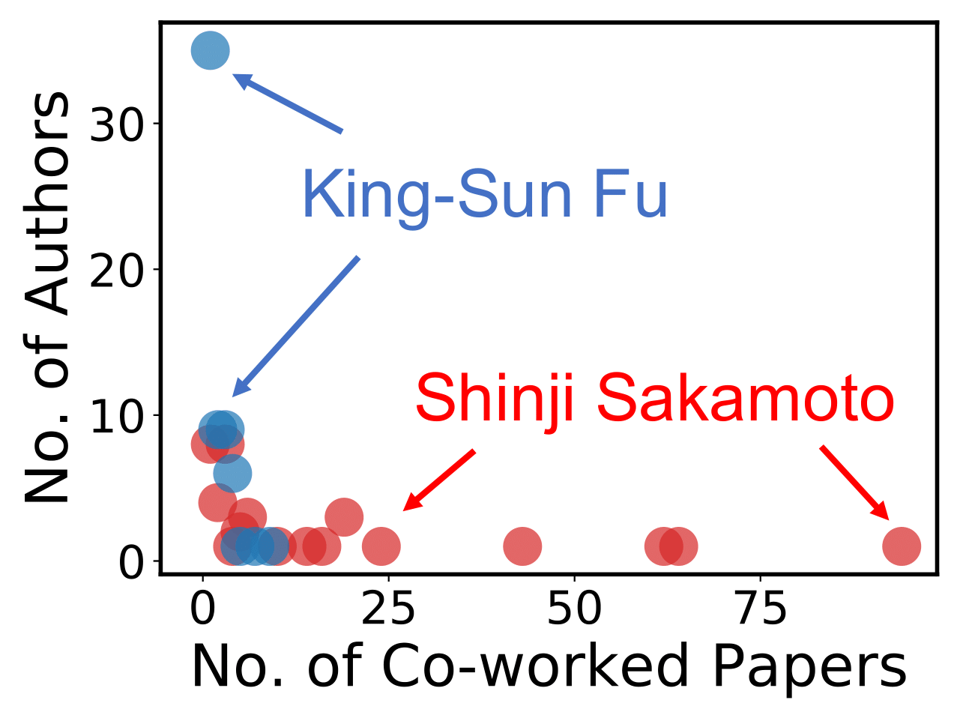}
		  \vspace{-5pt}
          \caption{Two researchers}
          \label{fig:case:A}
    \end{subfigure}
	\vspace{-5pt}
	\caption{Case studies on the \texttt{DBLP} dataset. Some authors deviate from the general pattern (\ref{fig:case:B}). Dr. Fu and Dr. Sakamoto differ in their co-working patterns (\ref{fig:case:A}). ~\label{fig:case:dblp}}
	\vspace{-10pt}
\end{figure}

\smallsection{Discoveries in Co-authorship Hypergraph:}
\texttt{DBLP} contains information of bibliographies of computer science publications. 
Each node represents an author, and each hyperedge consists of authors of a publication. 
The timestamp of the hyperedge is the year of publication. 
Here, we investigate how authors co-work with different researchers. 
For each author $v$ who have published at least 100 papers, we compute the average unexpectedness and burstiness scores of the hyperedges that $v$ is contained in, which we denote by $\mathsf{avg_U}(v)$ and $\mathsf{avg_B}(v)$, respectively. 
We analyze several authors whose ratio $\mathsf{avg_U}(v) / \mathsf{avg_B}(v)$ is the highest or the lowest.
Intuitively, authors with low ratios tend to co-work in a bursty manner with expected co-authors, while those with high ratios tend to work steadily with unexpected co-authors. Surprisingly, Dr. Bill Hancock, whose $\mathsf{avg_U}(v) / \mathsf{avg_B}(v)$ ratio is the lowest, published 186 papers all alone.
Furthermore, Dr. Hancock published 139 papers in 2000. On the other hand, Dr. Seymour Ginsburg, whose $\mathsf{avg_U}(v) / \mathsf{avg_B}(v)$ is the highest, published 114 papers from 1958 to 1999 (2.7 papers per year). In addition, 18 co-authors (out of 38) co-authored only one paper with Dr. Ginsburg. 
In fact, $\mathsf{avg_U}$ and $\mathsf{avg_B}$ of the most authors are clustered as seen in Figure~\ref{fig:case:B}, and the top authors are those with the largest (or the smallest) slope.
We further conduct case studies on two specific authors Dr. Shinji Sakamoto and Dr. King-Sun Fu, whose co-working patterns are very different. As seen in Figure~\ref{fig:case:A}, while Dr. Sakamoto collaborated on most papers with a few researchers, Dr. Fu enjoyed co-working with many new researchers. These findings support our intuition behind the proposed measures, $\mathsf{score_U}$ and $\mathsf{score_B}$.

\begin{figure}[t]
    \vspace{-2pt}
	\centering
	\includegraphics[width=0.6\linewidth]{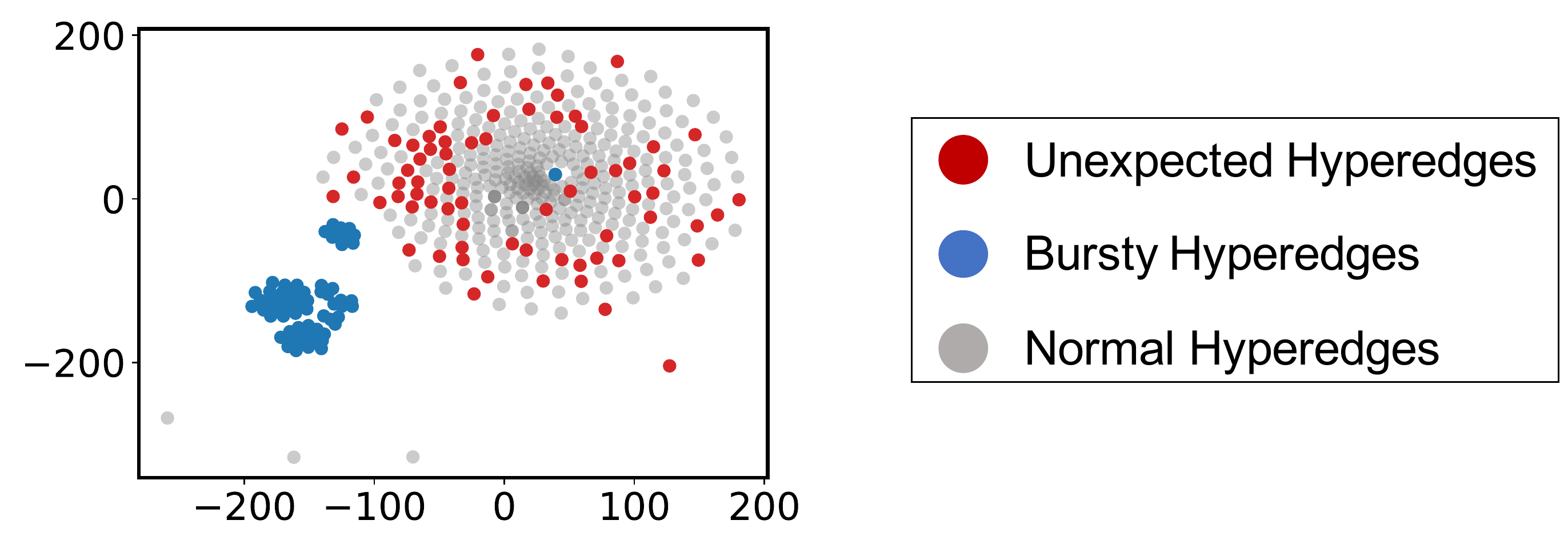}
	\vspace{-5pt}
	\caption{Case study on the \texttt{Cite-patent} dataset. Unexpected \& bursty hyperedges have different properties. \label{fig:case:patent}}
	\vspace{-10pt}
\end{figure}

\smallsection{Discoveries in Patent Citation Hypergraph}:
We use \texttt{Cite-patent}~\cite{tang2012patentminer}, which is a citation hypergraph where each node is a patent and each hyperedge is the set of patents cited by a patent.
The timestamp of each hyperedge is the year of the citation ranging from 2000 to 2012.
Using \method, we extract some hyperedges with high $\mathsf{score_U}$ or $\mathsf{score_B}$.
Then, we represent each hyperedge as a $|V|$-dimensional binary vector indicating which nodes belong to the hyperedge.
We visualize the hyperedges after reducing the dimension of the vectors via T-SNE in Figure~\ref{fig:case:patent}. While unexpected hyperedges are spread, bursty hyperedges are closely located, indicating structurally similar hyperedges arrive in bursts.
In addition, we closely examine the citation patterns of suspicious patents detected by \method. As seen in Figure~\ref{fig:crown:effectiveness}, patents with unexpected or bursty citations are effectively detected.

\smallsection{Discoveries in Online Q\&A Cite}:
We share 
the results of a case study using \texttt{Tags-overflow}.
In the dataset, nodes are tags and hyperedges are the set of tags attached to a question. 
Hyperedges with high $\mathsf{score_U}$ (i.e., sets of unexpected keywords) include: \{\textit{channel, ignore, antlr, hidden, whitespace}\}, \{\textit{sifr, glyph, stling, text-styling, embedding}\}, and \{\textit{retro-computing, boot, floppy, amiga}\}.
Hyperedges with high $\mathsf{score_B}$ (i.e., sets of bursty keywords) include: \{\textit{python, javascript}\}, \{\textit{java, adobe, javascript}\}, and \{\textit{c\#, java}\}. 
Notably, sets of unpopular tags tend to have high unexpectedness, while those containing popular keywords, such as \textit{python} and \textit{javascript}, have high burstiness.

\subsection{Q3. Scalability}
\label{sec:experiment:scalability}
To evaluate the scalability of \method, we measure how rapidly it updates the hypergraph summary and computes the anomaly scores as the number of hyperedges grows. To this end, we upscale \texttt{Email-Enron}, which originally consists of 10,885 hyperedges, by $2^1$ to $2^{17}$ times, and measure the total runtime of \method. As seen in Figure~\ref{fig:crown:scalability}, the total runtime is linear in the number of hyperedges, which is consistent with our theoretical analysis (Theorem~\ref{thm:total_time} in Section~\ref{sec:method}). 
That is, the time taken for processing each hyperedge is near constant. 
Notably, \method is scalable enough to process a stream of $1.4$ billion hyperedges within $2.5$ hours.

\subsection{Q4. Parameter Analysis}
\label{sec:experiment:paramter}
\begin{figure}
    	\centering
    	\begin{subfigure}[b]{.25\textwidth}
              \centering
              \includegraphics[width=0.995\columnwidth]{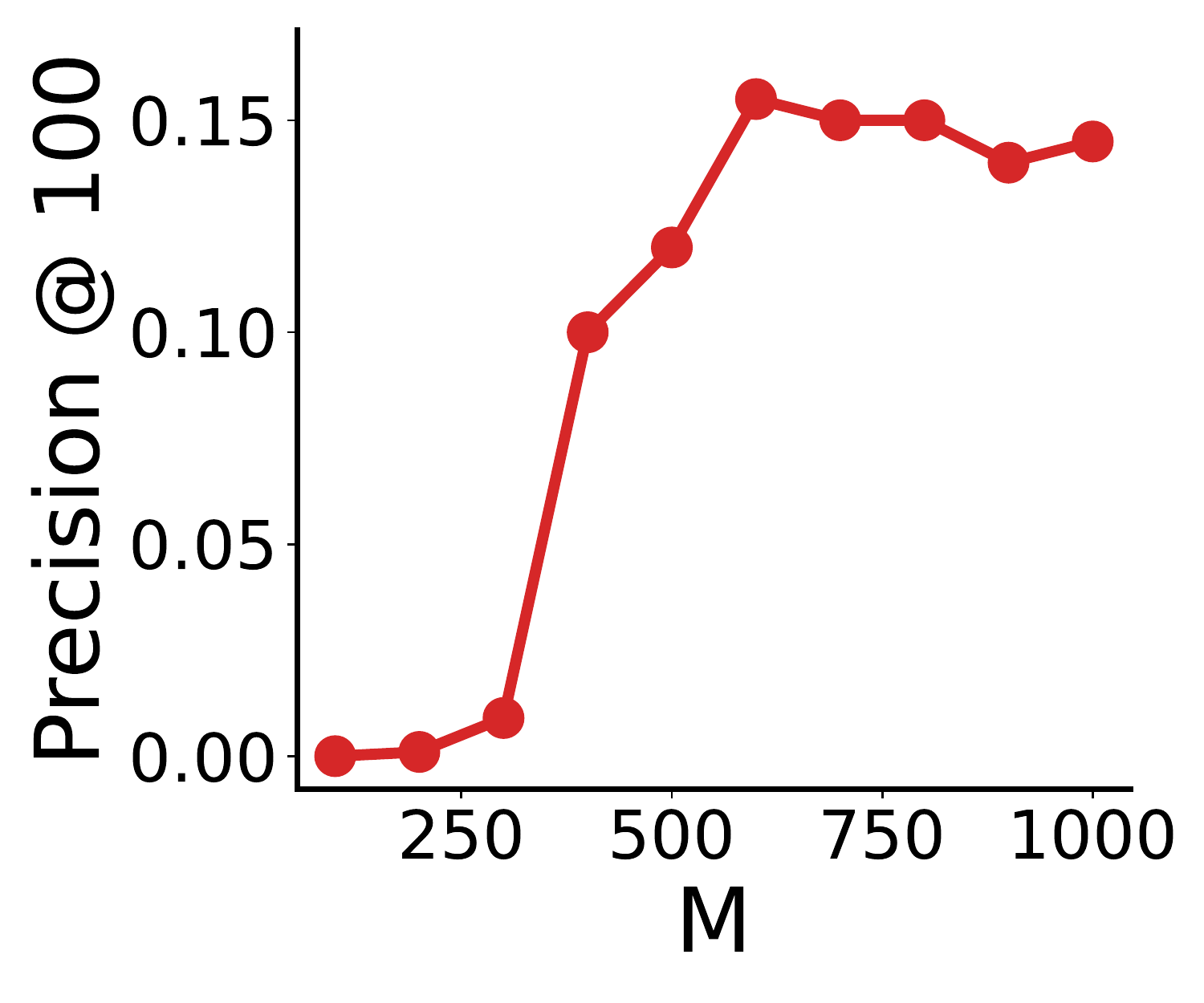}
    		  \vspace{-12pt}
              \caption{Effect of $M$}
              \label{fig:scalability:E}
        \end{subfigure}
        \hspace{-7pt}
        \begin{subfigure}[b]{.25\textwidth}
              \centering
              \includegraphics[width=0.995\columnwidth]{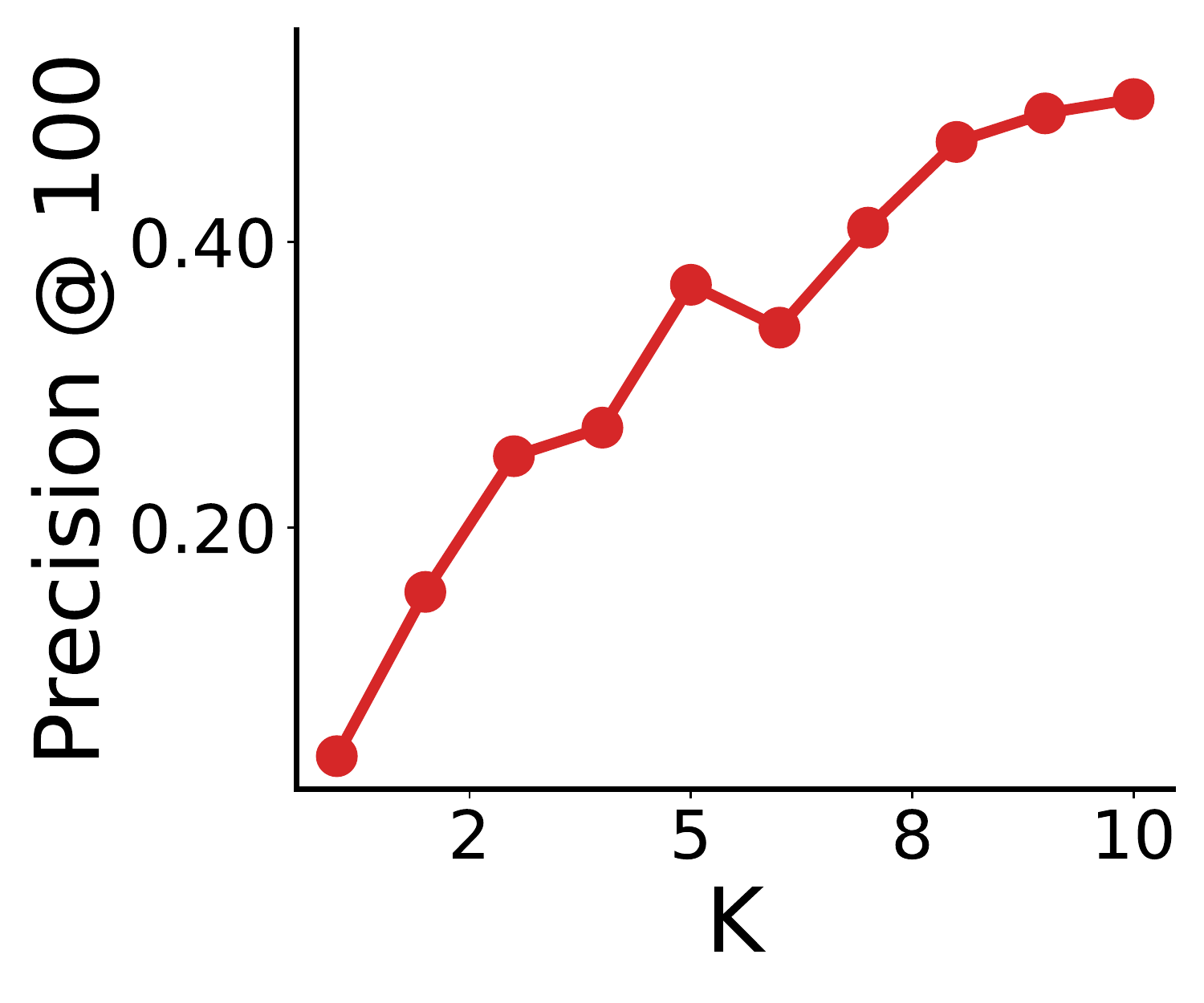}
    		  \vspace{-12pt}
              \caption{Effect of $K$}
              \label{fig:scalability:E}
        \end{subfigure}
        \hspace{-7pt}
        \begin{subfigure}[b]{.25\textwidth}
              \centering
              \includegraphics[width=0.995\columnwidth]{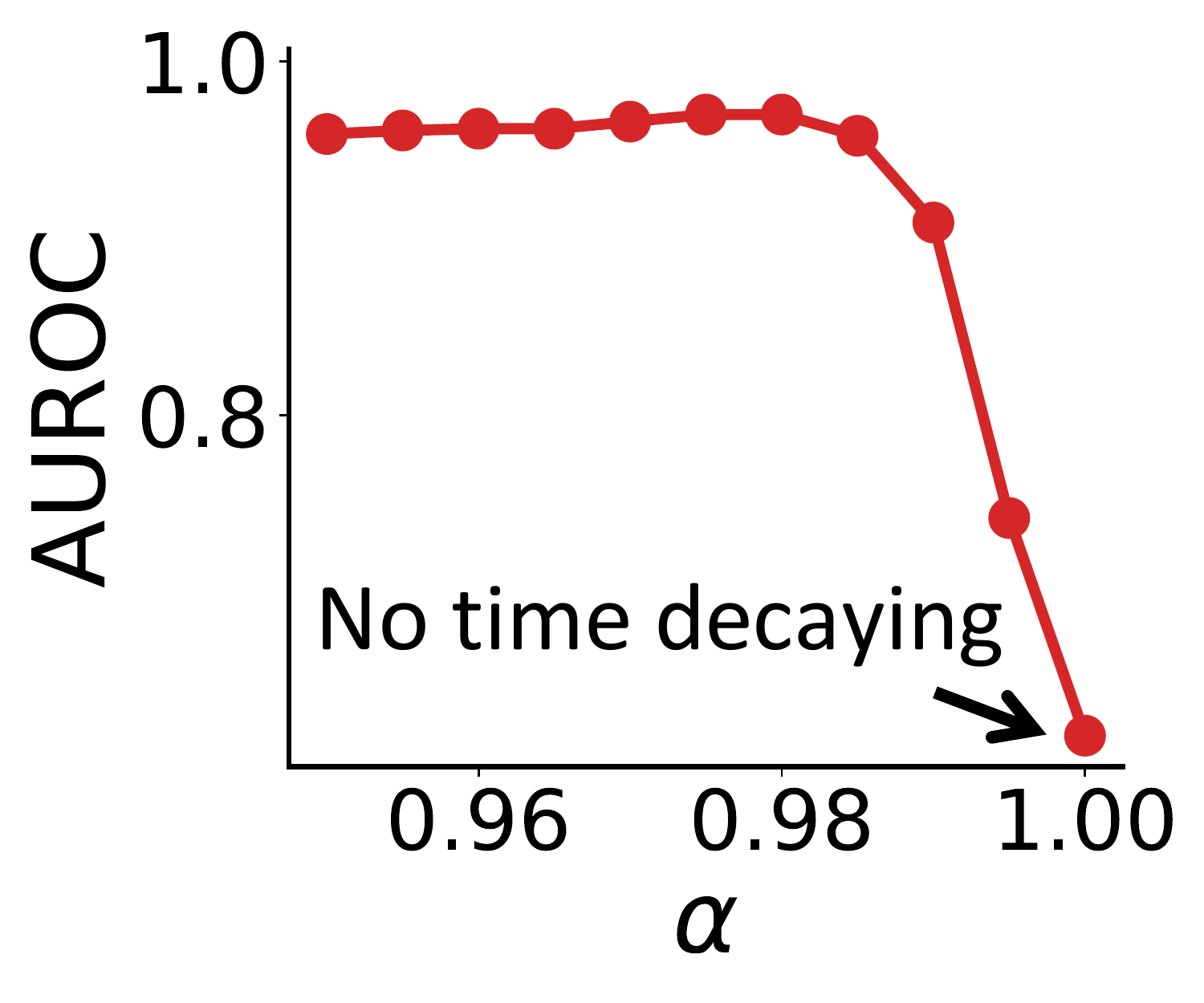}
    		  \vspace{-12pt}
              \caption{Effect of $\alpha$}
              \label{fig:scalability:E}
        \end{subfigure} \\
    	\vspace{-5pt}
    	\caption{The performance of \method depends on the number of supernodes ($M$), the number of hash functions ($K$), and time decaying parameter $\alpha$ in the \real dataset. \label{fig:parameter}}
    	\vspace{-10pt}
    \end{figure}

We evaluate \method under different parameter settings, and the results are shown in Figure~\ref{fig:parameter}. 
In most cases, there is a positive correlation with $M$ (i.e., the number of supernodes) and $K$ (i.e.,  the number of hash functions).
Intuitively, a larger number of supernodes and hash functions collectively reduce the variance due to randomness introduced by hash functions. 
However, since the space usage is dependent on these parameters,  a trade-off between the accuracy and the space usage should be considered. 
In addition, properly setting $\alpha$ (i.e., time decaying parameter) improves the accuracy, as shown in Figure~\ref{fig:parameter}, which indicates that not only structural information but also the temporal information is critical in detecting anomalies in hyperedge streams.

    \section{Conclusion}
    \label{sec:conclusion}
In this work, we propose \method, an online anomaly detector for hyperedge streams. \method maintains a random-walk-based hypergraph summary with constant space, and it is incrementally updated in near real-time. Using the summary, \method computes two anomaly scores that are effective in identifying (a) hyperedges composed of unexpected combinations of nodes and (b) those appearing in bursts.
Our experiments demonstrate the speed, accuracy, and effectiveness of \method in (semi-)real datasets.
The source code and datasets are publicly available at \url{https://github.com/geonlee0325/HashNWalk}.


   \smallsection{Acknowledgements:} This work was supported by National Research Foundation of Korea (NRF) grant funded by the Korea government (MSIT) (No. NRF-2020R1C1C1008296) and Institute of Information \& Communications Technology Planning \& Evaluation (IITP) grant funded by the Korea government (MSIT) (No. 2019-0-00075, Artificial Intelligence Graduate School Program (KAIST)).

\bibliographystyle{plain}
\bibliography{ref}

\begin{thebibliography}{10}

\bibitem{aggarwal2011outlier}
Charu~C Aggarwal, Yuchen Zhao, and S~Yu Philip.
\newblock Outlier detection in graph streams.
\newblock In {\em ICDE}, 2011.

\bibitem{akoglu2010oddball}
Leman Akoglu, Mary McGlohon, and Christos Faloutsos.
\newblock Oddball: Spotting anomalies in weighted graphs.
\newblock In {\em PAKDD}, 2010.

\bibitem{akoglu2015graph}
Leman Akoglu, Hanghang Tong, and Danai Koutra.
\newblock Graph based anomaly detection and description: a survey.
\newblock {\em DMKD}, 29(3):626--688, 2015.

\bibitem{bandyopadhyay2016topological}
Bortik Bandyopadhyay, David Fuhry, Aniket Chakrabarti, and Srinivasan
  Parthasarathy.
\newblock Topological graph sketching for incremental and scalable analytics.
\newblock In {\em CIKM}, 2016.

\bibitem{belth2020mining}
Caleb Belth, Xinyi Zheng, and Danai Koutra.
\newblock Mining persistent activity in continually evolving networks.
\newblock In {\em KDD}, 2020.

\bibitem{benson2018simplicial}
Austin~R Benson, Rediet Abebe, Michael~T Schaub, Ali Jadbabaie, and Jon
  Kleinberg.
\newblock Simplicial closure and higher-order link prediction.
\newblock {\em PNAS}, 115(48):E11221--E11230, 2018.

\bibitem{benson2018sequences}
Austin~R Benson, Ravi Kumar, and Andrew Tomkins.
\newblock Sequences of sets.
\newblock In {\em KDD}, 2018.

\bibitem{beutel2013copycatch}
Alex Beutel, Wanhong Xu, Venkatesan Guruswami, Christopher Palow, and Christos
  Faloutsos.
\newblock Copycatch: stopping group attacks by spotting lockstep behavior in
  social networks.
\newblock In {\em WWW}, 2013.

\bibitem{bhatia2020midas}
Siddharth Bhatia, Bryan Hooi, Minji Yoon, Kijung Shin, and Christos Faloutsos.
\newblock Midas: Microcluster-based detector of anomalies in edge streams.
\newblock In {\em AAAI}, 2020.

\bibitem{chakrabarti2004autopart}
Deepayan Chakrabarti.
\newblock Autopart: Parameter-free graph partitioning and outlier detection.
\newblock In {\em PKDD}, 2004.

\bibitem{chang2020f}
Yen-Yu Chang, Pan Li, Rok Sosic, MH~Afifi, Marco Schweighauser, and Jure
  Leskovec.
\newblock F-fade: Frequency factorization for anomaly detection in edge
  streams.
\newblock In {\em WSDM}, 2021.

\bibitem{chitra2019random}
Uthsav Chitra and Benjamin Raphael.
\newblock Random walks on hypergraphs with edge-dependent vertex weights.
\newblock In {\em ICML}, 2019.

\bibitem{choe2022midas}
Minyoung Choe, Jaemin Yoo, Geon Lee, Woonsung Baek, U~Kang, and Kijung Shin.
\newblock Midas: Representative sampling from real-world hypergraphs.
\newblock In {\em WWW}, 2022.

\bibitem{choo2022persistence}
Hyunjin Choo and Kijung Shin.
\newblock On the persistence of higher-order interactions in real-world
  hypergraphs.
\newblock In {\em SDM}, 2022.

\bibitem{dal2015calibrating}
Andrea Dal~Pozzolo, Olivier Caelen, Reid~A Johnson, and Gianluca Bontempi.
\newblock Calibrating probability with undersampling for unbalanced
  classification.
\newblock In {\em SSCI}, 2015.

\bibitem{ding2020more}
Kaize Ding, Jianling Wang, Jundong Li, Dingcheng Li, and Huan Liu.
\newblock Be more with less: Hypergraph attention networks for inductive text
  classification.
\newblock In {\em EMNLP}, 2020.

\bibitem{do2020structural}
Manh~Tuan Do, Se-eun Yoon, Bryan Hooi, and Kijung Shin.
\newblock Structural patterns and generative models of real-world hypergraphs.
\newblock In {\em KDD}, 2020.

\bibitem{eswaran2018sedanspot}
Dhivya Eswaran and Christos Faloutsos.
\newblock Sedanspot: Detecting anomalies in edge streams.
\newblock In {\em ICDM}, 2018.

\bibitem{eswaran2018spotlight}
Dhivya Eswaran, Christos Faloutsos, Sudipto Guha, and Nina Mishra.
\newblock Spotlight: Detecting anomalies in streaming graphs.
\newblock In {\em KDD}, 2018.

\bibitem{hayashi2020hypergraph}
Koby Hayashi, Sinan~G Aksoy, Cheong~Hee Park, and Haesun Park.
\newblock Hypergraph random walks, laplacians, and clustering.
\newblock In {\em CIKM}, 2020.

\bibitem{hooi2016fraudar}
Bryan Hooi, Hyun~Ah Song, Alex Beutel, Neil Shah, Kijung Shin, and Christos
  Faloutsos.
\newblock Fraudar: Bounding graph fraud in the face of camouflage.
\newblock In {\em KDD}, 2016.

\bibitem{huang2009video}
Yuchi Huang, Qingshan Liu, and Dimitris Metaxas.
\newblock Video object segmentation by hypergraph cut.
\newblock In {\em CVPR}, 2009.

\bibitem{hwang2008learning}
TaeHyun Hwang, Ze~Tian, Rui Kuangy, and Jean-Pierre Kocher.
\newblock Learning on weighted hypergraphs to integrate protein interactions
  and gene expressions for cancer outcome prediction.
\newblock In {\em ICDM}, 2008.

\bibitem{karypis1999multilevel}
George Karypis, Rajat Aggarwal, Vipin Kumar, and Shashi Shekhar.
\newblock Multilevel hypergraph partitioning: Applications in vlsi domain.
\newblock {\em TVLSI}, 7(1):69--79, 1999.

\bibitem{kim2020hypergraph}
Eun-Sol Kim, Woo~Young Kang, Kyoung-Woon On, Yu-Jung Heo, and Byoung-Tak Zhang.
\newblock Hypergraph attention networks for multimodal learning.
\newblock In {\em CVPR}, 2020.

\bibitem{kook2020evolution}
Yunbum Kook, Jihoon Ko, and Kijung Shin.
\newblock Evolution of real-world hypergraphs: Patterns and models without
  oracles.
\newblock In {\em ICDM}, 2020.

\bibitem{lee2021how}
Geon Lee, Minyoung Choe, and Kijung Shin.
\newblock How do hyperedges overlap in real-world hypergraphs? - patterns,
  measures, and generators.
\newblock In {\em WWW}, 2021.

\bibitem{lee2020hypergraph}
Geon Lee, Jihoon Ko, and Kijung Shin.
\newblock Hypergraph motifs: Concepts, algorithms, and discoveries.
\newblock {\em PVLDB}, 13(11):2256--2269, 2020.

\bibitem{lee2021thyme+}
Geon Lee and Kijung Shin.
\newblock Thyme+: Temporal hypergraph motifs and fast algorithms for exact
  counting.
\newblock In {\em ICDM}, 2021.

\bibitem{leontjeva2012fraud}
Anna Leontjeva, Konstantin Tretyakov, Jaak Vilo, and Taavi Tamkivi.
\newblock Fraud detection: Methods of analysis for hypergraph data.
\newblock In {\em ASONAM}, 2012.

\bibitem{mao2019multiobjective}
Mingsong Mao, Jie Lu, Jialin Han, and Guangquan Zhang.
\newblock Multiobjective e-commerce recommendations based on hypergraph
  ranking.
\newblock {\em Information Sciences}, 471:269--287, 2019.

\bibitem{park2009anomaly}
Youngser Park, C~Priebe, D~Marchette, and Abdou Youssef.
\newblock Anomaly detection using scan statistics on time series hypergraphs.
\newblock In {\em LACTS}, 2009.

\bibitem{ranshous2017efficient}
Stephen Ranshous, Mandar Chaudhary, and Nagiza~F Samatova.
\newblock Efficient outlier detection in hyperedge streams using minhash and
  locality-sensitive hashing.
\newblock In {\em Complex Networks}, 2017.

\bibitem{shin2016corescope}
Kijung Shin, Tina Eliassi-Rad, and Christos Faloutsos.
\newblock Patterns and anomalies in k-cores of real-world graphs with
  applications.
\newblock {\em KAIS}, 54(3):677--710, 2018.

\bibitem{shin2017densealert}
Kijung Shin, Bryan Hooi, Jisu Kim, and Christos Faloutsos.
\newblock Densealert: Incremental dense-subtensor detection in tensor streams.
\newblock In {\em KDD}, 2017.

\bibitem{silva2008hypergraph}
Jorge Silva and Rebecca Willett.
\newblock Hypergraph-based anomaly detection of high-dimensional
  co-occurrences.
\newblock {\em TPAMI}, 31(3):563--569, 2008.

\bibitem{tang2012patentminer}
Jie Tang, Bo~Wang, Yang Yang, Po~Hu, Yanting Zhao, Xinyu Yan, Bo~Gao, Minlie
  Huang, Peng Xu, Weichang Li, et~al.
\newblock Patentminer: topic-driven patent analysis and mining.
\newblock In {\em KDD}, 2012.

\bibitem{tang2016graph}
Nan Tang, Qing Chen, and Prasenjit Mitra.
\newblock Graph stream summarization: From big bang to big crunch.
\newblock In {\em SIGMOD}, 2016.

\bibitem{yang2019revisiting}
Dingqi Yang, Bingqing Qu, Jie Yang, and Philippe Cudre-Mauroux.
\newblock Revisiting user mobility and social relationships in lbsns: A
  hypergraph embedding approach.
\newblock In {\em WWW}, 2019.

\bibitem{yoon2019fast}
Minji Yoon, Bryan Hooi, Kijung Shin, and Christos Faloutsos.
\newblock Fast and accurate anomaly detection in dynamic graphs with a
  two-pronged approach.
\newblock In {\em KDD}, 2019.

\bibitem{yu2018netwalk}
Wenchao Yu, Wei Cheng, Charu~C Aggarwal, Kai Zhang, Haifeng Chen, and Wei Wang.
\newblock Netwalk: A flexible deep embedding approach for anomaly detection in
  dynamic networks.
\newblock In {\em KDD}, 2018.

\bibitem{zhao2011gsketch}
Peixiang Zhao, Charu~C Aggarwal, and Min Wang.
\newblock gsketch: On query estimation in graph streams.
\newblock {\em PVLDB}, 5(3):193--204, 2011.

\bibitem{zhou2007learning}
Dengyong Zhou, Jiayuan Huang, and Bernhard Sch{\"o}lkopf.
\newblock Learning with hypergraphs: Clustering, classification, and embedding.
\newblock In {\em NeurIPS}, 2007.

\end{thebibliography}

\end{document}